\documentclass{article}
\usepackage[utf8]{inputenc}
\usepackage[T1]{fontenc}
\usepackage{authblk}
\usepackage{setspace}
\usepackage[margin=1in]{geometry}
\usepackage{graphicx}
\graphicspath{{./figures/}}
\usepackage{subcaption}
\usepackage{amssymb}
\usepackage{amsmath}
\usepackage{enumitem}
\usepackage{siunitx}
\usepackage[dvipsnames]{xcolor}
\usepackage{csquotes}
\usepackage{lineno}
\usepackage[toc,page]{appendix}
\usepackage{multicol}
\usepackage[authoryear,round]{natbib}
\definecolor{DarkBurntOrange}{RGB}{200,80,0}
\usepackage{hyperref}
\hypersetup{
  colorlinks=true,
  urlcolor=ForestGreen,
  linkcolor=DarkBurntOrange,
  citecolor=blue
}
\title{Cosmological viability of anisotropic inflation in Thurston spacetimes}

\author[1]{Devika J.S. \thanks{devikaj20@iiserb.ac.in}}
\author[1]{Tanay Gupta \thanks{tanay23@iiserb.ac.in}}
\author[1]{Sukanta Panda \thanks{sukanta@iiserb.ac.in}}

\affil[1]{Department of Physics, Indian Institute of Science Education \& Research, Bhopal, India}

\date{}

\usepackage[authoryear]{natbib}
\usepackage{totcount}
\usepackage{etoolbox}

\newtotcounter{citcount}

\AtBeginEnvironment{thebibliography}{%
  \let\oldbibitem\bibitem
  \renewcommand{\bibitem}{\stepcounter{citcount}\oldbibitem}%
}

\begin{document}
\maketitle

\begin{abstract}
Abstract: Recent observations of large-scale statistical isotropy violations have prompted the adoption of anisotropic cosmological models that account for inherent directional curvature. Studies of these anisotropic spacetimes have shown how they can explain the evolutionary dynamics and light propagation in the universe. Here, we consider one such interesting set of spacetimes that preserve homogeneity but place no constraint on isotropy during the inflationary epoch, to examine whether we can address the possibility of anisotropic inflation in the universe. Researchers have proposed inflationary models in which a vector field coupled to the inflaton is found to violate the cosmic no-hair theorem for the anisotropic Bianchi type I spacetime, due to the existence of a stable anisotropically inflationary fixed point. Lately, this study has been extended to axisymmetric spacetimes of Bianchi type II, III, and the Kantowski-Sachs metric, and it has been inferred that the entire family of spacetimes is attracted to the anisotropic Bianchi I fixed point. By constructing inflationary models where the spatial slices are anisotropic Thurston 3-geometries, we demonstrate that the intrinsic eccentricity of the background geometry induces an isotropy-violating vector field. This field, through its coupling to the inflaton, triggers a secondary phase of anisotropic inflation. We perform dynamical stability and phase-space analyses to assess the feasibility of anisotropic inflation. The results for the considered set of Thurston geometries showed the presence of a unique, stable inflationary fixed point that converges, similar to those in Bianchi spacetimes, thereby indicating the cosmological viability of inflation with anisotropic hair.  
\end{abstract}

\newpage
\tableofcontents

%============================================================================================================
\newpage
\section{Introduction}
Understanding the evolution of the Universe and describing its underlying geometry are foundational aspects of modern cosmology. The study of the Universe's global topology and local geometry, governed by the Einstein field equations, provides critical insight into cosmic structure. In this framework, the spatial curvature—characterized as flat, positively curved, or negatively curved—is determined by the total stress-energy content within the spacetime, rather than mass-density alone.

The development of cosmological models can be traced from early geocentric and heliocentric frameworks to paradigm-shifting advances introduced by Einstein’s theory of gravity and Edwin Hubble’s discovery of the expanding Universe. These breakthroughs motivated the formulation of dynamical cosmological models, ultimately leading to Big Bang theory, inflationary cosmology, and the standard $\Lambda$CDM model.

The Standard $\Lambda$CDM (Cold Dark Matter) model, which represents the current best understanding of cosmic evolution and dynamics, is based on the Cosmological Principle. The principle is stated as
\begin{center}
\textit{Viewed on a sufficiently large scale, the properties of the universe are the same for all observers}
\end{center}
The principle elucidates the structural assertion of homogeneity and isotropy at large scales (larger than 100 Mpc) of the universe, which involves adopting the spatially homogeneous and isotropic Friedmann-Lemaître-Robertson-Walker (FLRW) metric (\cite{ellis1999cosmological}) to describe the universe's background spacetime.

The prominence of the standard inflationary $\Lambda$CDM paradigm stems from its consistency with the high-precision measurements of temperature fluctuations and polarization patterns in the Cosmic Microwave Background (CMB). Successive data from the COBE (\cite{smoot1992structure, bennett1994cosmic, smoot1999cobe}), WMAP (\cite{liu2009improved, souradeep2006angular, tegmark2003high, bennett2003microwave}), Planck (\cite{ade2016planck, n2020planck, schwarz2016cmb, lamarre2003planck, aghanim2020planck}) Boomerang (\cite{montroy2003measuring, gurzadyan2003ellipticity}), DASI (\cite{carlstrom2003status, leitch2002measuring}), ACT (\cite{aiola2020atacama}) and SPT (\cite{hou2014constraints}) missions, together with the explanation of accelerated expansion of the universe (\cite{riess1998observational, frieman2008dark}) and estimation of large scale structure properties by SDSS, DES \& DESI surveys (\cite{doroshkevich2004large, to2025dark, labini2025large}) and primordial light element abundances in the universe (\cite{alpher1948evolution, peebles1966primordial, wagoner1967synthesis, reeves1974origin, esmailzadeh1991primordial, cooke2014precision, cyburt2016big}) have allowed for the precise determination of its parameters, establishing it as a robust framework for describing the large-scale structure of the universe. Despite its efficacy, this model faces several discrepancies and anomalies in describing the full cosmology. These include the works \cite{copi2007uncorrelated, hansen2009power, peebles2025status, akarsu2024lambdacdm, sugiyama1994cosmic, scott1995microwave, koivisto2011possibility, wojtak2013orbital, bunn1996anisotropic, bartlett1999standard}. In particular, the works \cite{copi2010large, schwarz2016cmb, bunn2010large} highlight the CMB large angle anomalies, while those in \cite{de2004significance} and \cite{copi2004multipole} have focused on instances of violation in spatial isotropy from the anomalous alignment of the quadrupole and octopole of the CMB based on the WMAP data and also in indicating the existence of a preferred axis or direction of the universe corresponding to the CMB dipole. Furthermore, ambiguities related to the spatial curvature of the universe, as evidenced by disparities in the curvature density parameter reported in other prominent works, such as \cite{koivisto2011possibility}, make the accurate measurement and the existence of a directional dependence of the spatial curvature uncertain.

Given the limitations of existing cosmological models, we focus on more comprehensive models that can explain the universe's underlying spatial geometry and curvature, and efficiently account for its cosmological dynamics. Therefore, we look into the generalisation of a broader set of spacetimes by relaxing the assumption of spatial isotropy but retaining homogeneity, thus exploring homogeneous, anisotropic spacetimes through the context of Thurston geometries (\cite{thurston1982three, awwad2024large}).

The paper is categorized as follows: in section \ref{B}, we give a basic review of Thurston geometries along with a brief outline of cosmic inflation and its associated \textit{no-hair theorem}; in section \ref{inflationary} the basic pipelines for the inflationary model we'll be considering and the general set (Einstein + extended) of field equations; in section \ref{III} the determination of \textit{truly anisotropic} ones from the full set of proposed Thurston geometries along with their corresponding components to be used in our analyses; in sections \ref{IV} \& \ref{V} the methodology and procedure for the determination of stable points along with their results and in section \ref{VI} the re-formulation for the complete set of field equation s to probe any occurrence of anisotropic cosmic inflation. Finally, in section \ref{VII}, we present our analyses of observations \& results and in section \ref{VIII}, we conclude this work.  

%============================================================================================================
\section{Thurston spacetimes} \label{B}
The large-scale geometry of the universe breaks down into $\mathbb{R} \times \Sigma_3$ for $\mathbb{R}$ describing the cosmic time and $\Sigma_3$, the three-dimensional spatial manifold. Inducing anisotropy in $\Sigma_3$ can be modelled by representing its geometry as one of the eight geometries outlined in William Thurston's Geometrisation Conjecture (\cite{thurston1982three}), which was proven by Grigori Perelman (\cite{perelman2002entropy, perelman2003finite, perelman2003ricci})

\textbf{Thurston-Perelman’s geometrization theorem}:
\begin{center}
\textit{The interior of every compact 3-manifold has a canonical decomposition into pieces which have geometric structures}
\end{center}
This theorem defines an ordering within the set of geometries and concludes that any maximal, simply connected, three-dimensional geometry X that admits a compact quotient is equivalent to one of the eight geometries described in the conjecture (\cite{awwad2024large}).

We address the spacetimes developed in \cite{awwad2024large} under the framework of Thurston-Perelman's geometrisation theorem, which presents spacetime metrics with large-scale anisotropies in the spatial sections of the metric. Given below are the eight model Thurston geometries :
\begin{center}
    \begin{tabular}{>{\hspace{10pt}}l<{\hspace{10pt}} >{\hspace{10pt}}l<{\hspace{10pt}} >{\hspace{5pt}}l<{\hspace{5pt}} >{\hspace{10pt}}l<{\hspace{10pt}}}
i. $\mathbb{E}^3$ & iii. $S^3$ & v. $\mathbb{E}\times S^2$ & vii. $Nil$\\
ii. $\mathbb{H}^3$ & iv. $\mathbb{E}\times \mathbb{H}^2$ & vi. $\widetilde{U(\mathbb{H}^2)}$ & viii. $ Solv$\\
\end{tabular}\\
\end{center}
A general and explicit metric representation of the Thurston spacetimes is given by :
\begin{equation}
ds^2 = -dt^2 + a^2(t) \left[\gamma_{ij,\Sigma_3(Thurston)} dx^i dx^j\right]
\end{equation}
\vspace{20 pt}
where a(t) is the scale factor and $\gamma_{ij,\Sigma_3(Thurston)}$ is specific to each of the Thurston geometries, as shown below:
\begin{itemize}
  \item \textbf{$\mathbb{E}^3$, $\mathbb{H}^3$ and $S^3$}\\
They represent the exact spatial slices from the FLRW spacetime and, therefore, are the only isotropic geometries within this framework, with $S^3$ being a special case of twisted $S^2 \times S^1$ (\cite{weeks2001shape}). Under a hyperspherical coordinate system, they can be described as
\begin{equation}
    ds^2 = -dt^2 +a^2(t)[d\chi^2 + S_\kappa^2(\chi)d\Omega^2] = -dt^2 +a^2(t)[d\chi^2 + S_k^2(\chi)(d\theta^2+sin^2 \theta d\phi^2)],\label{eq:FRW m}
\end{equation}
\begin{equation} \label{rule}
    S_\kappa (\chi) =
    \begin{cases}
    \frac{\text{sin}(\chi \sqrt{\kappa})}{\sqrt{\kappa}}, & \kappa > 0\\
    \chi, & \kappa = 0\\
    \frac{\text{sinh}(\chi \sqrt{-\kappa})}{\sqrt{-\kappa}}, & \kappa < 0
    \end{cases} 
\end{equation}
\vspace{20 pt}
where $\kappa$ represents the curvature parameter and $\chi \in [0, \infty)$, $\theta \in [0, \pi)$ and $\phi \in [0, 2\pi)$ respectively.\\
  \item \textbf{$\mathbb{E}\times \mathbb{H}^2$ and $\mathbb{E}\times S^2$}\\\begin{equation}
        ds^2 = -dt^2 +a^2(t)[d\chi^2 + S_\kappa^2(\chi)d\phi^2 + dz^2 ].\label{eq:RH2 m}
\end{equation}
where $S_\kappa(\chi)$ is given by equation \eqref{rule}. Note that this is a cylindrical geometry and the flat case ($\kappa$ = 0) is not allowed in this spacetime to distinguish it from the isotropic geometry $\mathbb{E}^3$.
  \item \textbf{$\widetilde{U(\mathbb{H}^2)}$ }\\\begin{equation}
       ds^2 = -dt^2 +a^2(t)[dx^2 + \cosh^2(x\sqrt{-\kappa})dy^2 + (dz + \sinh(x\sqrt{-\kappa})dy)^2].\label{eq:UH2 m}\vspace{20 pt}
\end{equation}
\item \textbf{Nil}\\\begin{equation}
       ds^2 = -dt^2 +a^2(t)[dx^2 + (1-\kappa x^2)dy^2 + dz^2 -2x\sqrt{-\kappa} dy dz].\label{eq:Nil m}\vspace{20 pt}
\end{equation}
\item \textbf{Solv}\\
\begin{equation}
      ds^2 = -dt^2 +a^2(t)[e^{2z\sqrt{-\kappa}}dx^2 + e^{-2z\sqrt{-\kappa}}dy^2 + dz^2].\label{eq:Solv m}\vspace{20 pt}
\end{equation}
\end{itemize}
The large-scale geometry of the universe's spatial sections can be depicted as a mosaic, composed of one or more of the eight Thurston geometries, seamlessly joined together. However, because the radius of curvature of any of these considered geometries exceeds the Hubble radius today, we assume for simplicity that the observable spatial sections of the universe can be accounted for by just one of these Thurston geometries, as asserted in \cite{awwad2024large}.

The rapid expansion of the universe, captured during the period of cosmic inflation, explains how it dilutes the irregularities and asymmetries from the early universe, thereby smoothing it out onto a homogeneous, isotropic background at large scales. The conventional inflationary models that account for the observed flatness as well as uniformity of the universe without any special initial conditions are demonstrated as a consequence of the  \textit{cosmic no-hair theorem} (\cite{kitada199207579cosmic}), which asserts that any initial anisotropies and inhomogeneities in the spatial geometry decay rapidly during inflation.

Recently, inflationary models that violate isotropy and thereby the no-hair conjecture have been studied to elucidate the theoretical possibility of \textit{inflation with anisotropic hair}. In ref. \cite{Watanabe_2009}, spatially flat Bianchi type I models with vector field coupled to the inflaton were studied and were observed to contradict the no-hair theorem by proving the possibility of anisotropic inflation via the existence of a stable anisotropically inflationary fixed point. Furthermore, they validated this possibility in \cite{watanabe2011imprints} by studying the statistical anisotropy from anisotropic inflation, including tensor perturbations, and examining how it was imprinted in the two-point correlations of CMB temperature fluctuations and polarisations. Lately, research has focused on introducing curvature into the Bianchi I background geometry, considering axisymmetric spacetimes with three-dimensional rotational invariance for Bianchi type II, III, and Kantowski-Sachs metrics (cited in \cite{Hervik_2011}). For exponential potentials specified for the scalar field and for the coupling function between the scalar and vector fields, it was analyzed that the anisotropic Bianchi type I fixed point is an attractor for the entire family of these spacetimes, which concludes the cosmological viability of inflation with anisotropic hair. Here, we perform a similar study of inflationary modelling and dynamical stability analysis using Thurston spacetimes to determine whether they exhibit a stable anisotropically inflationary fixed point, thereby validating anisotropic inflation.

%============================================================================================================
\section{The inflationary model}\label{inflationary}
We refer to the inflationary model considered for the axisymmetric spacetimes from \cite{Hervik_2011} described by the action term S for the metric $g_{\mu \nu}$, scalar field $\phi$ and vector field $A_\mu$:
\begin{equation} \label{action}
S = \int d^4x \, \sqrt{-g} \left[ \frac{M_{(Pl)}^2}{2} R - \frac{1}{2} (\nabla \phi)^2 - V(\phi) 
- \frac{1}{4} f^2(\phi) F_{\mu\nu} F^{\mu\nu} \right].
\end{equation}
The kinetic term of the scalar field is given by \((\nabla \phi)^2 = g^{\mu\nu} \nabla_{\mu} \phi \nabla_{\nu} \phi\)\quad and \( F_{\mu\nu} = \nabla_{\mu} A_{\nu} - \nabla_{\nu} A_{\mu} \) represents the field strength of the vector field. $g$ is the determinant of the corresponding metric, $R$ is the Ricci scalar and $M_{(Pl)}$ the reduced Planck mass.\\
Following the same formalism as in \cite{Hervik_2011}, we consider the scalar field potential V($\phi$) and the coupling function f($\phi$) to be exponential types:
\begin{equation} \label{expoten}
V(\phi) = V_0 e^{\lambda \frac{\phi}{M_{(Pl)}}},
\end{equation}
\begin{equation}
f(\phi) = f_0 e^{Q \frac{\phi}{M_{(Pl)}}},
\end{equation}
where $\lambda$ and $Q$ are constant parameters characterizing the model. The assumption of slow-roll inflation implies $\lambda <<1$ and the coupling constant $Q$ is a free parameter, while for considering a positive potential, we constrain $V_0 > 0$. This formalism results in the following set of equations of motion, for variation with respect to $g_{\mu \nu}$, $\phi$ and $A_\mu$ :
\begin{equation}
M_{(Pl)}^2 G_{\mu\nu} = T_{\mu\nu}^{(\phi)} + T_{\mu\nu}^{(A)},\label{FEq1}
\end{equation}
\begin{equation}
\nabla_\mu T^{\mu(\phi)}_\nu = -Q \frac{2\mathcal{L}_A}{M_{(Pl)}} \nabla_\nu \phi,\label{phiTmunu}
\end{equation}
\begin{equation}
\nabla_\mu T^{\mu(A)}_\nu = +Q \frac{2\mathcal{L}_A}{M_{(Pl)}} \nabla_\nu \phi\label{ATmunu},
\end{equation}
where $T^{\mu(\phi)}_\nu$ and $T^{\mu(A)}_\nu$ are the energy-momentum tensors for the scalar and the vector fields respectively, and $\mathcal{L}_A$ the vector field Lagrangian such that $\mathcal{L}_A = -\frac{1}{4} f^2(\phi) F^{\mu\nu} F_{\mu\nu}$.\\
From equations \eqref{phiTmunu} and \eqref{ATmunu}, we note the conservation of energy-momentum tensor, stated by $\nabla_\mu \left( T^{\mu (\phi)}_{\nu} + T^{\mu (A)}_{\nu} \right) = 0$, i.e. the coupling between the scalar and vector fields parametrized by the coupling constant Q results in the transfer of energy and momentum among these fields. The shear degree of freedom associated with the anisotropy in our spacetimes is regarded as sourced by the axisymmetric vector field $A_\mu$, which in this case is coupled to the inflaton.

The energy-momentum tensor for the scalar and vector fields is given by
\begin{equation}
T_{\nu}^{\mu (\phi)} = -\delta_{\nu}^{\mu} \left( \frac{1}{2} (\nabla \phi)^2 + V(\phi) \right) 
+ \nabla^{\mu} \phi \nabla_{\nu} \phi,
\end{equation}
\begin{equation}
T_{\nu}^{\mu (A)} = -f^2(\phi) F^{\mu \alpha} F_{\alpha \nu} 
- \frac{1}{4} f^2(\phi) \delta_{\nu}^{\mu} F_{\alpha\beta} F^{\alpha\beta}
\end{equation}
where $\delta^\mu_\nu = g^{\mu\alpha} g_{\alpha\nu}$.

%============================================================================================================
\section{On Thurston geometries} \label{III}
Understanding the set of geometries discussed within Thurston's framework and the inherent anisotropies present in them, as noted even in other prominent works such as \cite{graham2010observing} and \cite{smith2025cosmological}, we assess and refine them to fit within the inflationary model developed above. From ref. \cite{awwad2024large}, studying the evolutionary dynamics of the cosmological background in Thurston spacetimes for isotropic scale factors along all spatial directions includes the derivation of corresponding Einstein tensors of each spacetime. This gives the fluid solutions compatible with the Einstein tensors with an additional shear tensor term when compared with the FLRW case, to address the anisotropies in them:
\begin{equation}
T^\mu_\nu = (\rho +p) u^\mu u_\nu + p\delta^\mu_\nu +\pi^\mu_\nu
\end{equation}
where $\rho$(t) denotes the energy density and p(t) the pressure of the fluid, $u^\mu$(t) is the fluid velocity vector and $\pi^\mu_\nu$ corresponds to the shear tensor ($\pi^\mu_\nu$ is symmetric, traceless, transverse and vanishes for a perfect fluid \textit{ansatz} in the case of standard FLRW geometry). Hence, these generalised fluid solutions will be subjected to shear constraints (\cite{awwad2024large}), which can be resolved by the introduction of anisotropic scale factors along different spatial directions.

Introducing anisotropic expansion drops off the shear tensor by splitting the single isotropic scale factor $a(t)$ into three independent components $a_1(t)$, $a_2(t)$ and $a_3(t)$. Considering the perfect fluid \textit{ansatz}, we can constrain these different scale factors through cancelling out their off-diagonal components of Einstein tensors for the respective spacetimes:
\begin{equation}
ds^2 = -dt^2 + (\gamma_{ij,\Sigma_3(Thurston)} a_i(t) a_j(t) dx^i dx^j)
\end{equation}
\begin{table}[h]
    \centering
    \setlength{\arrayrulewidth}{0.35mm}
    \setlength{\tabcolsep}{20pt}
    \renewcommand{\arraystretch}{1.5}
    \begin{tabular}{ |c|c| } 
     \hline
      Spacetime & Scale Factor Constraints\\
     \hline
       $\mathbb{E}^3$/$\mathbb{H}^3$/$S^3$  & \(a_1=a_2=a_3=a\) \\
     \hline
     $\mathbb{E}\times \mathbb{H}^2$/$\mathbb{E} \times S^2$ & \(a_1=a_2=a,a_3=b\)\\
     \hline
     \( \widetilde{U(\mathbb{H}^2)} \) & \(a_1=a_2=a_3=a\)\\
     \hline
      $Nil$ & \(a_1= b, a_2=a_3=a\)\\
     \hline
     $Solv$ & \(a_1=a_2=a, a_3=b\)\\ 
     \hline
    \end{tabular}
    \caption{Scale factor constraints for anisotropic expansion of Thurston spacetimes}
    \label{tab:scalefactor_constraints}
\end{table}
Interpreting the scale factor constraints in table \ref{tab:scalefactor_constraints}, we note that for the \(\widetilde{U(\mathbb{H}^2)} \) geometry, all the anisotropic scale factors of expansions are equal, implying that anisotropic expansion cannot dissolve its shear tensor contribution into this geometry. Further, since $\mathbb{E}^3$, $\mathbb{H}^3$ and $S^3$ are already isotropic geometries recognised under the FLRW framework, we therefore discontinue our anisotropic studies on these geometries and focus on the four anisotropic geometries, namely, $\mathbb{E}\times \mathbb{H}^2$ , $\mathbb{E} \times S^2$, $Nil$ and $Solv$.

We present the metric equations for these four spacetimes subjected to their constraints on the anisotropic scale factors from table \ref{tab:scalefactor_constraints}. \\
\begin{itemize}
    \item \textbf{$\mathbb{E}\times \mathbb{H}^2$ and $\mathbb{E}\times S^2$}
    \begin{equation} \label{rh2}
          ds^2 = -dt^2 + a^2(t)[d\chi^2 + S_k^2(\chi)d\phi^2] + b^2(t)dz^2
\end{equation}
\item \textbf{Nil}
\begin{equation} \label{nil}
         ds^2 = -dt^2 + a^2(t)[(1-\kappa x^2)dy^2+dz^2-2x\sqrt{-\kappa}dydz] + b^2(t)dx^2
\end{equation}
\item \textbf{Solv}
\begin{equation} \label{solv}
         ds^2 = -dt^2 + a^2(t)[e^{2z\sqrt{-\kappa}}dx^2 + e^{-2z\sqrt{-\kappa}}dy^2] + b^2(t)dz^2
\end{equation}\\
\end{itemize}
From equations \eqref{rh2}, \eqref{nil} and \eqref{solv}, we analyze that the plane spanned by the spatial directions mapped to the scale factor $a(t)$ will have a rotational symmetry and the spatial direction corresponding to $b(t)$ will be orthogonal to this plane of rotational symmetry i.e., it becomes the axis of symmetry. So, in adherence with the previous discussions, we would fix the vector field sourcing the shear parallel to this axis of symmetry, which would be $z$ axis for $\mathbb{E}\times \mathbb{H}^2$, $\mathbb{E}\times S^2$ and $Solv$ geometries, and $x$ axis for $Nil$ geometry. Therefore, if we define $a_{||}$ and $a_\perp$ as the scale factors in the direction parallel and perpendicular to the vector field $A_\mu$, respectively, then the anisotropic scale factors can be redefined as 
\begin{subequations}
    \begin{gather}
        a_{||} =b (t)= e^{\alpha(t)}e^{-2\sigma(t)}\\
        a_\perp = a(t)= e^{\alpha(t)}e^{\sigma(t)}
    \end{gather}
\end{subequations}
where $\alpha(t)$ and $\sigma(t)$ are functions of time. This will transform the corresponding Hubble parameters like
\begin{subequations}
    \begin{gather}
         H_{||} =H_b= \dot{\alpha}-2\dot{\sigma}\\
        H_\perp = H_a= \dot{\alpha}+\dot{\sigma}
    \end{gather}\label{HeqRH2}
\end{subequations}
and the average Hubble rate H would become
\begin{equation}
    H = \frac{2}{3}H_a + \frac{1}{3}H _b = \dot{\alpha}\label{alpha eq}
\end{equation}
For the inflationary model, we quantify the acceleration in terms of the deceleration parameter \textit{q}
\begin{equation}
q = -1 - \frac{\dot{H}}{H^2},\label{q para}
\end{equation}
where $q<0$ implies acceleration, which is equivalent to
\begin{equation}
\frac{d^2}{dt^2} e^\alpha > 0
\end{equation}
such that \( e^\alpha \) conforms with the isotropic scale factor defined in terms of the geometric mean of the anisotropic scale factors, respectively.

The redefinition of the anisotropic scale factors gives a generalised expression for the considered set of spacetimes given by:
\begin{equation}
     ds^2 = -dt^2 + e^{2\alpha(t)}[e^{-4\sigma(t)}w^1\otimes w^1 +e^{2\sigma(t)}w^2\otimes w^2+e^{2\sigma(t)}w^3\otimes w^3].\label{gen metric}
\end{equation}
We consider a single electric-type vector field $A_\mu$ aligned with the axis of symmetry of each geometry. Therefore, the field strength tensor for the $\mathbb{E}\times \mathbb{H}^2$, $\mathbb{E}\times S^2$ and $Solv$ geometries is given by: 
\begin{equation}
F = \frac{1}{2} F_{\mu\nu}(t) \, dx^\mu \wedge dx^\nu = F_{30}(t) \, dz\wedge dt.
\end{equation}
In the gauge $A_0=0$, the vector potential would be:
 \begin{equation}
A = A_\mu dx^\mu = A_z(t) dz,
\end{equation}
and hence \begin{equation}
F_{30} = \nabla_zA_0 -\nabla_0A_z =-\dot{A}_z.
\end{equation}
The field strength tensor for the $Nil$ geometry is given by: 
\begin{equation}
F = \frac{1}{2} F_{\mu\nu}(t) \, dx^\mu \wedge dx^\nu = F_{10}(t) \, dx \wedge dt.
\end{equation}
 In the gauge $A_0=0$, the vector potential would be:
 \begin{equation}
A = A_\mu dx^\mu = A_x(t) dx,
\end{equation}
and therefore \begin{equation}
F_{10} = \nabla_xA_0 -\nabla_0A_x =-\dot{A}_x.
\end{equation}
We introduce two spacetime-related coefficients $r_1$ and $r_2$ from
\begin{equation}
^3\dot{R} = -2(\dot{\alpha} + r_1 \dot{\sigma})\quad^3R\label{Ricci}
\end{equation}
for $^3R$ defined as the three-dimensional Ricci scalar of constant time hypersurfaces and $r_2$ determining the coupling strength between curvature and energy density of the vector field.

Considering the matter sources, the generalized vector field strength would be expressed as
\begin{equation}
F = \dot{v}(t) \, w^0 \wedge w^1 + (b+2r_2 \kappa v(t))\, w^2 \wedge w^3,
\end{equation}
where $v(t)$ is a dynamical degree of freedom and b is a constant. The field is homogeneous, axisymmetric, and includes both electric- and magnetic-type components. With reference to \cite{Hervik_2011}, we restrict the magnetic field component by setting $b=0$ such that now the $r_2$ value determines the existence of the magnetic component. However, for the geometries $\mathbb{E}\times \mathbb{H}^2$ , $\mathbb{E} \times S^2$, $Nil$ and $Solv$ discussed here, there is no coupling between the spatial curvature and the energy density of the vector field ($r_2=0$), which implies the absence of magnetic-type component.

Therefore,
\begin{equation}
    F = \dot{v} \, w^0 \wedge w^1
\end{equation}
and vector potential,
\begin{equation}
    \vec{A}= vdz
\end{equation}
for $\mathbb{E}\times \mathbb{H}^2$, $\mathbb{E}\times S^2$ and $Solv$ geometries but
\begin{equation}
    \vec{A}= vdx
\end{equation}
for $Nil$ geometry.\\
The components of the energy-momentum tensor are written as:
\begin{equation} \label{A}
T^\mu_\nu =
    \begin{cases}
    \text{diag} (-\rho,p_\perp,p_\perp,p_{||}), \hspace{0.5cm} \mathbb{E}\times \mathbb{H}^2/S^2 \hspace{0.2cm} \& \hspace{0.2cm} Solv\\
    \text{diag} (-\rho,p_{||},p_\perp,p_\perp), \hspace{0.5cm} Nil
    \end{cases} 
\end{equation}
which are physical quantities representing the energy density and pressure measured in the fluid's rest frame.

The energy density and pressure contributions from scalar and vector fields are given by
\begin{subequations}
    \begin{gather}
        \rho = \rho_\phi +\rho_A,\\
        p_{||} = p_\phi +(p_A)_{||},\\
         p_{\perp} = p_\phi +(p_A)_\perp,
    \end{gather}
\end{subequations}
such that
\begin{subequations}
    \begin{gather}
        \rho_\phi = \frac{1}{2} \dot{\phi}^2 + V(\phi) \label{40a}\\
p_\phi = \frac{1}{2} \dot{\phi}^2 - V(\phi)\\
\rho_A = f^2 e^{-2\alpha + 4\sigma} 
\left( \frac{1}{2} \dot{v}^2 - r_2 (^3R) v^2 \right) = \frac{1}{2} f^2 e^{-2\alpha + 4\sigma} 
\dot{v}^2
    \end{gather}
\end{subequations}
From the assumptions followed from \cite{Hervik_2011}, the equation  of state for inflationary expansion of the spacetimes will be constrained to
\begin{subequations}
    \begin{gather}
        (p_A)_{||} = -\rho_A , \label{41a} \\(p_A)_\perp = +\rho_A. \label{41b}
    \end{gather}
\end{subequations}

%============================================================================================================
\section{Field equations and dynamic system formulation} \label{IV}
We shall now consider the set of Thurston spacetimes generalised by the metric equation \eqref{gen metric} and possessing inherent anisotropic curvature with rotational symmetry, which produces a shear as represented by the vector field aligned parallel to the $w^1$ direction. From the spacetime metrics in equations \eqref{rh2}, \eqref{nil} and \eqref{solv}, we calculate the Ricci scalar corresponding to each of the spacetimes to obtain the three-dimensional Ricci scalar of constant time hypersurfaces and thereby, find the value of constant $r_1$ from equation \eqref{Ricci} as given in table \ref{tab:Geom}.
\begin{table}[h]
    \centering
    \setlength{\arrayrulewidth}{0.35mm}
    \setlength{\tabcolsep}{20pt}
    \renewcommand{\arraystretch}{1.5}
    \begin{tabular}{ |c|c|c|} 
     \hline
      Geometries & $^3R $& $r_1$ \\
     \hline
      $\mathbb{E}\times \mathbb{H}^2 / S^2$ & $ ^3R = 2\kappa e^{-2\alpha-2\sigma}$  & 1\\
     \hline
      $Nil$ & $^3R = \frac{\kappa}{2} e^{-2\alpha+4\sigma}$ & -2 \\
     \hline
      $Solv$ &$ ^3R = 2\kappa e^{-2\alpha+4\sigma}$  & -2\\
     \hline
    \end{tabular}
    \caption{Geometric variables of Thurston spacetimes}
    \label{tab:Geom}
\end{table}
\newline
Now, we consider each of these Thurston spacetimes, derive their field equations, and formulate them as dynamical systems parametrised by a set of dimensionless variables that represent the system in phase space. Furthermore, we rewrite the field equations in terms of these dimensionless variables, yielding a set of autonomous first-order differential equations that describe the dynamics of these variables in the phase space.

%============================================================================================================
\subsection{\textbf{$\mathbb{E}\times \mathbb{H}^2 / S^2$ geometries}} \label{RH2 eq sec}
The field equations \eqref{FEq1} - \eqref{ATmunu} obtained from the action term of the model would be expressed as given below for the corresponding Einstein tensor and energy-momentum tensor components of the spacetime:
\begin{equation} \label{H^2 for RH}
    H^2 - \dot{\sigma}^2 = \frac{\rho}{3M_{(Pl)}^2} - \frac{^3R}{6}
\end{equation}
\begin{equation}
    \ddot{\sigma} + 3H\dot{\sigma} = \frac{p_\perp - p_{||}}{3M_{(Pl)}^2} - \frac{^3R}{6},
\end{equation}
\begin{equation}
    \dot{H} + 3H^2 = \frac{1}{2M_{(Pl)}^2} \left( \rho - \frac{p_{||}}{3}  - \frac{2 p_\perp}{3} \right) - \frac{^3R}{3},
\end{equation}
    \begin{equation}
^3\dot{R} = -2 (\dot{\alpha} + \dot{\sigma}) ^3R,
\end{equation}
\begin{equation}
    \dot{\rho}_\phi + 3H (\rho_\phi + p_\phi) = Q \mathcal{L}_A \frac{2\dot{\phi}}{M_{(Pl)}},
\end{equation}
\begin{equation}
\dot{\rho}_A + 4\rho_A(H + \dot{\sigma})  = -Q \mathcal{L}_A \frac{2\dot{\phi}}{M_{(Pl)}},
\end{equation}
for $\mathcal{L}_A = \frac{1}{2} f^2(\phi)\dot{v}^2 e^{-2\alpha  + 4\sigma}$.\\
Subsequently, we propose a set of dimensionless variables $\{X, Y, Z, \Omega_\kappa\}$ and represent the field equations above in terms of these variables.\\
We introduce the shear variable, X
\begin{equation}
     X = \frac{H_\perp - H}{H} = \frac{\dot{\sigma}}{\dot{\alpha}}\\
    \quad \quad .\label{dynvarinfl}
\end{equation}
the curvature variable $\Omega_\kappa$
\begin{equation}
    \Omega_\kappa = -\frac{^3R}{6H^2}
\end{equation}
and the other two variables
\begin{equation}
     Y = \frac{\dot{\phi}}{M_{(Pl)} H},
\end{equation}
and
\begin{equation}
    Z = \frac{f e^{-\alpha + 2\sigma} \dot{v}}{M_{(Pl)} H}.
\end{equation}
respectively. It is to be noted that while Z is proportional to $\dot{v}$ that contains $\sqrt{-g}$ which brings inhomogeneity in the case of $\mathbb{E}\times \mathbb{H}^2 / S^2$ geometries, one can easily prove using variable separation analysis the removal of spatial inhomogeneities requires the corresponding term to be rendered unity, thereby ensuring that the resulting set of equations uphold the fundamental assumption of spatial homogeneity. We skip that here, and use $\alpha$ in equation \eqref{alpha eq} as the time parameter to perform the differentiation such that:
\begin{equation}
    X' = \frac{dX}{d\alpha}= \frac{\dot{X}}{H},
\end{equation}
and the acceleration equation  of the spacetime expressed in terms of the phase space variables will be:
\begin{equation}
    \frac{H'}{H} = \frac{-Y^2}{2}-\frac{Z^2}{3}-\Omega_\kappa-3X^2
\end{equation}
This will subsequently help us derive the autonomous set of differential equations:
\begin{subequations}
\begin{gather}
        X' = \frac{1}{3}Z^2(X+1) + \Omega_\kappa(X+1)+X \left[3(X^2-1)+\frac{1}{2}Y^2 \right],\\
        Y' = (Y+\lambda)\left[\frac{1}{2}Y^2+3(X^2-1)\right]+\frac{1}{3}YZ^2+ \left(Q+\frac{\lambda}{2} \right)Z^2+\Omega_\kappa(3\lambda+Y),\\
        Z'= Z \left[3(X^2-1)+\frac{1}{2}Y^2-QY+1-2X+\frac{1}{3}Z^2 \right]+\Omega_\kappa Z,\\
        \Omega_\kappa'=2\Omega_\kappa \left[-1 -X+3X^2+\frac{Y^2}{2}+\frac{Z^2}{3}+\Omega_\kappa \right].
\end{gather}
\end{subequations}
Since we considered a positive potential for the scalar field, the dynamical variables of the system will be under the following constraint equation 
\begin{equation}
    X^2+\frac{Y^2}{6}+\frac{Z^2}{6}+\Omega_\kappa<1\label{constraintRinf}
\end{equation}
This bounds the range of the variables X in $(-1,1)$, Y and Z into $(-\sqrt{6},\sqrt{6)}$ and $\Omega_\kappa$ in $(0,1)$. However, for the $\mathbb{E}\times \mathbb{H}^2$ spacetime, $\Omega_\kappa<0$ due to negative curvature, which makes the other variables arbitrarily large. We also define various cosmological parameters of the spacetimes in terms of these dynamical variables. This includes the energy density parameters of scalar and vector fields, $\Omega_\phi$ and $\Omega_A$ respectively:
\begin{equation}
    \Omega_A=\frac{\rho_A}{3M_{(Pl)}^2H^2}=\frac{Z^2}{6},
\end{equation}
\begin{equation}
    \Omega_\phi=\frac{\rho_\phi}{3M_{(Pl)}^2H^2}=1-X^2-\Omega_\kappa-\frac{Z^2}{6}.
\end{equation}
For $\Omega_\phi=\Omega_{K.E}+\Omega_V$, the kinetic energy contributes from $\Omega_{K.E} = Y^2 / 6$ and equation \eqref{H^2 for RH} will be modified as
\begin{equation}
    X^2+\Omega_{K.E}+\Omega_{V}+\Omega_{A}+\Omega_{\kappa}=1.
\end{equation}\\
The deceleration parameter explained in equation \eqref{q para} becomes:
\begin{equation}
    q =  -1 + 3X^2+\frac{Y^2}{2}+\frac{Z^2}{3}+\Omega_\kappa.
\end{equation}

%============================================================================================================
\subsection{\textbf{Nil geometry}}
The field equations for the Nil spacetime would be written as:
\begin{equation}
 H^2 - \dot{\sigma}^2 = \frac{\rho}{3M_{(Pl)}^2} - \frac{^3R}{6},\label{H^2 for Nil}
\end{equation}
\begin{equation}
\ddot{\sigma} + 3H\dot{\sigma} = \frac{p_\perp - p_{||}}{3M_{(Pl)}^2} + 2\frac{^3R}{3},
\end{equation}
\begin{equation}
\dot{H} + 3H^2 = \frac{1}{2M_{(Pl)}^2} \left( \rho - \frac{p_{||}}{3}  - \frac{2 p_\perp}{3} \right) - 2\frac{^3R}{3},
\end{equation}
\begin{equation}
^3\dot{R} = -2 (\dot{\alpha} -2\dot{\sigma}) ^3R,
\end{equation}
\begin{equation}
\dot{\rho}_\phi + 3H (\rho_\phi + p_\phi) = Q \mathcal{L}_A \frac{2\dot{\phi}}{M_{(Pl)}},
\end{equation}
\begin{equation}
\dot{\rho}_A + 4 \rho_A(H + \dot{\sigma}) = -Q \mathcal{L}_A \frac{2\dot{\phi}}{M_{(Pl)}},
\end{equation}
where $\mathcal{L}_A = \frac{1}{2} f^2(\phi) \dot{v}^2e^{-2\alpha  + 4\sigma}$.\\
The set of dynamical variables defining the phase space of the system $\{X,Y,Z,\Omega_\kappa\}$ is the same as in section \ref{RH2 eq sec}.

The acceleration equation  of the Nil spacetime is modified as:
\begin{equation}
    \frac{H'}{H} = -\frac{Y^2}{2}-\frac{Z^2}{3}-\Omega_\kappa-2\Omega_\kappa X-3X^2
\end{equation}
This will subsequently help us derive the autonomous set of differential equations:
\begin{subequations}
\begin{gather}
        X'=-3X + \frac{Z^2}{3} -4\Omega\kappa -X\left(-3X^2-\frac{Y^2}{2}-\frac{Z^2}{3}-\Omega_\kappa-2X\Omega_\kappa\right)\label{Nil 1}\\
        Y' =QZ^2 -3Y +\frac{\lambda}{2}Y^2-3\lambda+3X^2\lambda+3\Omega_\kappa \lambda+\frac{\lambda Z^2}{2}-Y \left(\frac{-Y^2}{2}-\frac{Z^2}{3}-\Omega_\kappa-2X\Omega_\kappa-3X^2 \right)\\
        Z'= Z(-2QY-1 -4X)+Z(-1+2X+QY)-Z \left(\frac{-Y^2}{2}-\frac{Z^2}{3}-\Omega_\kappa-2X\Omega_\kappa-3X^2 \right)\\
        \Omega_\kappa'=\Omega_\kappa(-2 +4X)-2\Omega_\kappa \left(\frac{-Y^2}{2}-\frac{Z^2}{3}-\Omega_\kappa-2X\Omega\kappa-3X^2 \right)\label{Nil 4}
\end{gather}
\end{subequations}
The constraint equation is subjected to the positive potential of the scalar field is
\begin{equation}
    X^2+\frac{Y^2}{6}+\frac{Z^2}{6}+\Omega_\kappa<1.\label{constraintNilinf}
\end{equation}
and the deceleration parameter will be
\begin{equation}
    q =  -1 +3X^2+\frac{Y^2}{2}+\frac{Z^2}{3}+\Omega_\kappa+2X\Omega_\kappa.\label{deccel Nil}
\end{equation}

%============================================================================================================
\subsection{\textbf{Solv geometry}}
For the Solv spacetime, deriving the field equations from the corresponding tensor components formulates:
\begin{equation}
 H^2 - \dot{\sigma}^2 = \frac{\rho}{3M_{(Pl)}^2} - \frac{^3R}{6},\label{H^2 for Solv}
\end{equation}
\begin{equation}
\ddot{\sigma} + 3H\dot{\sigma} = \frac{p_\perp - p_{||}}{3M_{(Pl)}^2} +\frac{^3R}{3},
\end{equation}
\begin{equation}
\dot{H} + 3H^2 = \frac{1}{2M_{(Pl)}^2} \left( \rho - \frac{p_{||}}{3}  - \frac{2 p_\perp}{3} \right) - \frac{^3R}{3},
\end{equation}
\begin{equation}
 ^3\dot{R} = -2(\dot{\alpha}-2\dot{\sigma})^3R,
\end{equation}
\begin{equation}
\dot{\rho}_\phi + 3H (\rho_\phi + p_\phi) = Q \mathcal{L}_A \frac{2\dot{\phi}}{M_{(Pl)}},
\end{equation}
\begin{equation}
\dot{\rho}_A + 4\rho_A(H + \dot{\sigma})  = -Q \mathcal{L}_A \frac{2\dot{\phi}}{M_{(Pl)}},
\end{equation}
where $\mathcal{L}_A = \frac{1}{2} f^2(\phi) \dot{v}^2e^{-2\alpha  + 4\sigma}$.

Following the same definition for the set of phase space variables $\{X,Y,Z,\Omega_\kappa\}$  as in the previous sections, we derive the autonomous set of first-order differential equations that explains the system dynamics.
 
The acceleration equation  of the Solv spacetime will be:
\begin{equation}
    \frac{H'}{H} = -\frac{Y^2}{2}-\frac{Z^2}{3}-\Omega_\kappa-3X^2
\end{equation}
Proceeding with this:
\begin{subequations}
\begin{gather}
        X' =-3X + 3X^3+\frac{XY^2}{2}+\frac{Z^2}{3}+\frac{XZ^2}{3} -2\Omega\kappa +X\Omega_\kappa\label{Solv 1}\\
        Y' =QZ^2 -3Y +\frac{\lambda}{2}Y^2-3\lambda+3X^2\lambda+3\Omega_\kappa \lambda+\frac{\lambda Z^2}{2}-Y \left(\frac{-Y^2}{2}-\frac{Z^2}{3}-\Omega_\kappa-3X^2 \right)\\
        Z'= Z(-2QY-1 -4X)+Z(-1+2X+QY)-Z \left(\frac{-Y^2}{2}-\frac{Z^2}{3}-\Omega_\kappa-3X^2 \right)\\
        \Omega_\kappa'=2\Omega_\kappa(-1 +2X)-2\Omega_\kappa \left(\frac{-Y^2}{2}-\frac{Z^2}{3}-\Omega_\kappa-3X^2 \right).\label{Solv 4}
\end{gather}
\end{subequations}
Assuming a positive potential for the scalar field of the model, the constraint equation  is given by: 
\begin{equation}
    X^2+\frac{Y^2}{6}+\frac{Z^2}{6}+\Omega_\kappa<1,\label{constraintSolvinf}
\end{equation}
and the deceleration parameter will be:
\begin{equation}
    q =  -1 +3X^2+\frac{Y^2}{2}+\frac{Z^2}{3}+\Omega_\kappa.
\end{equation}

%============================================================================================================
\section{Dynamical stability analysis} \label{V}
The objective of this section is to investigate the phase-space structure and stability by applying the dynamical systems approach for the anisotropic inflationary models proposed through Thurston spacetimes. Understanding the field equations associated with each spacetime and formulating them as an autonomous system of differential equations, parametrised by the appropriate dynamical variables, helps describe the system's phase space. Furthermore, we identify the system's fixed points and classify them using linear stability theory (c.f. \cite{Bahamonde_2018}) to investigate the existence of stable, unique anisotropic fixed points, which would imply the cosmological feasibility of anisotropic inflation in Thurston spacetimes.

The fixed points (also called \textit{critical points}) are those points where the system remains in a steady state (or equilibrium) indefinitely. The critical points for each system are found by solving its corresponding set of differential equations and setting the equations to zero. The obtained points are then analysed to determine the system's stability in a linear regime near them.

The linear stability theory approximates the system linearly around the fixed points, $\frac{d\delta X^i}{d\alpha}=J\delta X^i$, and the eigenvalues of J, the \textit{Jacobian matrix} (or \textit{stability matrix}) evaluated at the critical points, convey the stability at that point. If all the eigenvalues of $J$ at a given critical point have positive real parts, then trajectories are repelled from that critical point, and it is an \textbf{unstable} point (or a \textbf{repeller}). If all eigenvalues have negative real parts, the point would attract all the nearby trajectories and it is  \textbf{stable} (\textbf{attractor}). Lastly, if at least two eigenvalues have real parts with opposite signs, the corresponding critical point is known as a \textbf{saddle point}, attracting trajectories in certain directions while repelling them in others.\\  

%============================================================================================================
\subsection{\textbf{$\mathbb{E}\times \mathbb{H}^2 / S^2$ geometries}}
In view of the assumptions made in section \ref{inflationary} which we restrain $\lambda>0$ and $\lambda<<1$, we found the critical points of the $\mathbb{E}\times \mathbb{H}^2 / S^2$ system as given in table \ref{tab:InflationRH2FP}, satisfying the constraint equation of the model equation \eqref{constraintRinf}. In table \ref{tab:InflationRH2EV} we give an outlook on the stability properties and deceleration parameters for the critical points obtained.
\begin{table}[h]
    \centering
    \setlength{\arrayrulewidth}{0.35mm}
    \setlength{\tabcolsep}{18pt}
    \renewcommand{\arraystretch}{1.5}
    \begin{tabular}{ |c|c|c|c|c| } 
     \hline
      Critical points & \( X\) & \( Y\) & \( Z^2\) & \(\Omega_\kappa\)\\
     \hline
      A & $\frac{\lambda Q-2}{3 Q^2}$ &$ \frac{-2}{Q}$ &$ \frac{6\lambda Q-12}{2Q^2}$ & 0 \\
     \hline
      B & 0 & $-\lambda$ & 0 & 0\\
     \hline
      C & $-1$ & 0 & $\frac{18\lambda}{2Q-\lambda}$ & $\frac{-3(2Q+\lambda)}{2Q-\lambda}$\\
     \hline
      D & $\frac{-2 + \lambda^2}{2(1+\lambda^2)}$ & $\frac{-3\lambda}{(1+\lambda^2)}$ & 0 & $\frac{3(-4 + \lambda^4)}{4(1+\lambda^2)^2} $\\
     \hline
      E & free & $\pm\sqrt{6-6X^2}$ & 0 & 0\\
     \hline
     F & $\frac{1}{2}$ & 0 & 0 & $\frac{3}{4}$\\
     \hline
      G &$ \frac{2+3Q\sqrt{2+4Q^2}}{4+6Q^2}$& $\frac{3Q-3\sqrt{(2+4Q^2)}}{2+3Q^2}$ & 0 & 0 \\
     \hline
     H &$ \frac{2- 3Q\sqrt{2+4Q^2}}{4+6Q^2}$& $\frac{3Q+3\sqrt{(2+4Q^2)}}{2+3Q^2}$ & 0 & 0 \\
     \hline
    \end{tabular}
    \caption{The valid critical points for the $\mathbb{E}\times \mathbb{H}^2 / S^2$ inflationary models}
    \label{tab:InflationRH2FP}
\end{table}
\begin{table}[h]
    \centering
    \setlength{\arrayrulewidth}{0.35mm}
    \setlength{\tabcolsep}{18pt}
    \renewcommand{\arraystretch}{1.5}
    \begin{tabular}{ |c|c|c|c|} 
     \hline
      Critical points & Existence & Stability & q \\
     \hline
      A & $ Q>>1$  & stable &  $~\left(-1+\frac{\lambda}{Q} \right)$\\
     \hline
      B & $- $ & stable/saddle  & $\left(-1 +\frac{\lambda^2}{2} \right)$ \\
     \hline
      C &$ \frac{\lambda}{Q}<2 $  & saddle &  -1\\
     \hline
      D & $- $   & saddle  & -1 +$\frac{3\lambda^2}{(2+2\lambda^2)}$\\
     \hline
      E &  $-$  &  unstable & 2\\
     \hline
     F &  $-$  &  saddle & $\frac{1}{2} $\\
     \hline
     G &  $-$  &  unstable &$2$\\
     \hline
      H &  $-$  &  unstable &$2$\\
     \hline
    \end{tabular}
    \caption{Stability analysis and deceleration parameters for the corresponding critical points of $\mathbb{E}\times \mathbb{H}^2$ and $\mathbb{E} \times S^2$ inflationary models}
    \label{tab:InflationRH2EV}
\end{table}\\
The evaluation of the system gives us four critical points A-D valid under the constraint equation \eqref{constraintRinf} from the positive potential consideration $(V(\phi)>0)$ and the remaining four points E-H, which are on the boundary, i.e. $V(\phi)=0$. Linear stability analysis given in appendix \ref{app:RH2 app} infers the presence of a unique, stable anisotropic attractor A, which exists only when $Q>>1$. The other points are examined to be either saddle or unstable from their eigenvalues, except for B, which is stable in a regime where the attractor A does not exist, thereby preserving its uniqueness. The value of the deceleration parameter q characterizes the points A-D as inflationary fixed points since $q<0$ and E-H as decelerating fixed points as $q>0$.

%============================================================================================================
\subsection{Nil geometry}
Finding and analysing the valid fixed points of the \textit{Nil} model with respect to equation \eqref{constraintNilinf} corresponds to the results in tables \ref{tab:InflationNilFP} and \ref{tab:InflationNilEV}.\\
\begin{table}[!h]
    \centering
    \setlength{\arrayrulewidth}{0.35mm}
    \setlength{\tabcolsep}{16pt}
    \renewcommand{\arraystretch}{1.5}
    \begin{tabular}{ |c|c|c|c|c| } 
     \hline
      Critical points & \( X\) & \( Y\) & \( Z^2\) & \(\Omega_\kappa\)\\
     \hline
      A & $\frac{\lambda Q-2}{3 Q^2}$ &$ \frac{-2}{Q}$ &$ \frac{6\lambda Q-12}{2Q^2}$ & 0 \\
     \hline
      B & 0 & $-\lambda$ & 0 & 0\\
     \hline
      C & $\frac{2Q - \lambda}{4(Q + \lambda)}$ & $-\frac{3}{Q + \lambda}$ & $-\frac{9}{Q^2} + \frac{27\lambda}{8Q}$ & $-\frac{3}{8}+\frac{33\lambda}{32Q}$\\
     \hline
      D & free & $\pm\sqrt{6-6X^2}$ & 0 & 0\\
     \hline
     E &$ \frac{2+3Q\sqrt{2+4Q^2}}{4+6Q^2}$& $\frac{3Q-3\sqrt{(2+4Q^2)}}{2+3Q^2}$ & 0 & 0 \\
     \hline
     F &$ \frac{2- 3Q\sqrt{2+4Q^2}}{4+6Q^2}$& $\frac{3Q+3\sqrt{(2+4Q^2)}}{2+3Q^2}$ & 0 & 0 \\
     \hline
      G & -1& 0 & 0 & 0 \\
      \hline
       H & 2& 0 & 0 & -3 \\
      \hline
       I & $-\frac{1+3Q^2}{4+3Q^2}$& $\frac{9Q}{4+3Q^2}$ &$ \frac{54}{(4+3Q^2)^2} $& $\frac{3}{8+6Q^2}$ \\
      \hline
    \end{tabular}
    \caption{The valid critical points for the Nil inflationary model}
    \label{tab:InflationNilFP}
\end{table}
\begin{table}[!h]
    \centering
    \setlength{\arrayrulewidth}{0.35mm}
    \setlength{\tabcolsep}{18pt}
    \renewcommand{\arraystretch}{1.5}
    \begin{tabular}{ |c|c|c|c|} 
     \hline
      Critical points & Existence & Stability & q \\
     \hline
      A & $ Q>>1$  & stable &  $~ \left(-1+\frac{\lambda}{Q} \right)$\\
     \hline
      B & $- $ & stable/saddle & $ \left(-1 +\frac{\lambda^2}{2} \right)$ \\
     \hline
      C &  $Q>>1$  &  saddle & $-1 + \frac{3\lambda}{2Q}$\\
     \hline
     D &  $-$  & unstable & 2\\
     \hline
     E &  $-$  &  unstable &$2$\\
     \hline
      F &  $-$  &  unstable &$2$\\
     \hline
      G&  $-$  &  unstable &$2$\\
     \hline
      H &  $-$  &  stable &$-4$\\
     \hline
     I &  $-$  &  saddle &$2-\frac{2}{Q^2}$\\
     \hline
    \end{tabular}
    \caption{Stability analysis and deceleration parameters for the corresponding critical points of the Nil inflationary model}
    \label{tab:InflationNilEV}
\end{table}
Solving the system of equations \eqref{Nil 1} - \eqref{Nil 4}, which pertains to the inflationary model developed from the $Nil$ spacetime and constraining the obtained critical points under equation \eqref{constraintNilinf}, resulted in a total of nine fixed points, out of which six points, D-I, are boundary points that satisfy equation \eqref{constraintNilinf} when it becomes equal to 1. Critical point A represents the inflationary, unique, and stable attractor of the system, which exists only in the regime of $Q>>1$. B presents the same scenario as in $\mathbb{E}\times \mathbb{H}^2 / S^2$ models, where it is stable in the non-existence regime of A and saddle otherwise. The deceleration parameter from equation \eqref{deccel Nil}, analysed at each of the fixed points, concludes that points A, B, C and H are inflationary while others are decelerating. We urge the reader to kindly refer to appendix \ref{app:Nil app} for a detailed explanation and analysis of each critical point of the \textit{Nil} inflationary model.

%============================================================================================================
\subsection{Solv geometry}
Finding the valid fixed points with respect to equation \eqref{constraintSolvinf} corresponds to the results in table \ref{tab:InflationSolvFP} and an outlook on to their properties discussed in table \ref{tab:InflationSolvEV}.
\begin{table}[!h]
    \centering
    \setlength{\arrayrulewidth}{0.38mm}
    \setlength{\tabcolsep}{16pt}
    \renewcommand{\arraystretch}{1.8}
    \begin{tabular}{ |c|c|c|c|c| } 
     \hline
      Critical points & \( X\) & \( Y\) & \( Z^2\) & \(\Omega_\kappa\)\\
     \hline
      A & $\frac{\lambda Q-2}{3 Q^2}$ &$ \frac{-2}{Q}$ &$ \frac{6\lambda Q-12}{2Q^2}$ & 0 \\
     \hline
      B & 0 & $-\lambda$ & 0 & 0\\
     \hline
      C & $\frac{2Q - \lambda}{4(Q + \lambda)}$ &$-\frac{3}{Q + \lambda}$&$\frac{-9}{Q^2} + \frac{9\lambda}{2Q}$& $-\frac{3}{4}+\frac{9\lambda}{4Q}$\\
     \hline
      D & $\frac{2 - \lambda^2}{4 + \lambda^2}$ & $-\left(\frac{6\lambda}{4 + \lambda^2}\right)
$ & 0 & $\frac{6(-2 + \lambda^2)}{(4 + \lambda^2)^2}
 $\\
     \hline
      E & free & $\pm\sqrt{6-6X^2}$ & 0 & 0\\
     \hline
     F & -1 & 0 & 0 & 0 \\
     \hline
     G & $-1 + \frac{3}{2Q^2}$ & $\frac{3}{Q}$ &$ \frac{9}{Q^4} $& $\frac{3}{2Q^2}$ \\
     \hline
     H &$ \frac{2+3Q\sqrt{2+4Q^2}}{4+6Q^2}$& $\frac{3Q-3\sqrt{(2+4Q^2)}}{2+3Q^2}$ & 0 & 0 \\
     \hline
     I &$ \frac{2- 3Q\sqrt{2+4Q^2}}{4+6Q^2}$& $\frac{3Q+3\sqrt{(2+4Q^2)}}{2+3Q^2}$ & 0 & 0 \\
     \hline
    \end{tabular}
    \caption{The valid critical points for the Solv inflationary model}
    \label{tab:InflationSolvFP}
\end{table}
\begin{table}[!h]
    \centering
    \setlength{\arrayrulewidth}{0.35mm}
    \setlength{\tabcolsep}{18pt}
    \renewcommand{\arraystretch}{1.5}
    \begin{tabular}{ |c|c|c|c|} 
     \hline
      Critical points & Existence & Stability & q \\
     \hline
      A & $ Q>>1$  & stable &  $~ \left(-1+\frac{\lambda}{Q} \right)$\\
     \hline
      B & $- $ & stable/saddle  & $ \left(-1 +\frac{\lambda^2}{2} \right)$ \\
     \hline
      C &$ Q>>1$  & saddle &   $~ \left(-1+\frac{3\lambda}{2Q} \right)$\\
     \hline
      D & $- $   & saddle  & -1\\
     \hline
      E &  $-$  &  unstable & 2\\
     \hline
     F &  $-$  & unstable& 2\\
     \hline
     G &  $-$  &  saddle &$2-\frac{3}{Q^2}$\\
     \hline
     H &  $-$  &  unstable &$2$\\
     \hline
      I &  $-$  &  unstable &$2$\\
     \hline
    \end{tabular}
    \caption{Stability analysis and deceleration parameters for the corresponding critical points of the Solv inflationary model}
    \label{tab:InflationSolvEV}
\end{table}\\
The fixed point analysis on the \textit{Solv} inflationary model given in equations \eqref{Solv 1} - \eqref{Solv 4} presents A-D as the valid critical points under the constraint equation \eqref{constraintSolvinf} and E-I as the boundary points. The positive energy condition imposed on the vector field limits the existence of the stable attractor A, and the saddle point C. B possesses the same stability criterion exhibited by other spacetimes, while the other critical points would either be saddle or unstable (refer to appendix \ref{app:Solv app}). Fixed points A-D are inflationary, while E-I are decelerating.

%============================================================================================================
\subsection{Phase flow analysis} \label{Phase flow analysis}
Studying the set of Thurston spacetimes by formulating them as anisotropic inflationary models helped us identify a unique anisotropic attractor in all the spacetimes represented by the critical point A. In addition to A we also have the fix-point B, which is stable in the parameter region where A does not exists, and is saddle in the region of existence of A. Therefore, the critical point B can also be the unique attractor in the phase space in case of peculiar initial conditions (such as $Z = 0$), which is the case for an initially vanishing vector field. However, for arbitrary initial conditions, the inflationary systems described by any of these Thurston spacetimes converge towards the fixed point A.

\begin{figure}
\begin{subfigure}{.5\textwidth}
    \centering
    \includegraphics[width=0.65\linewidth]{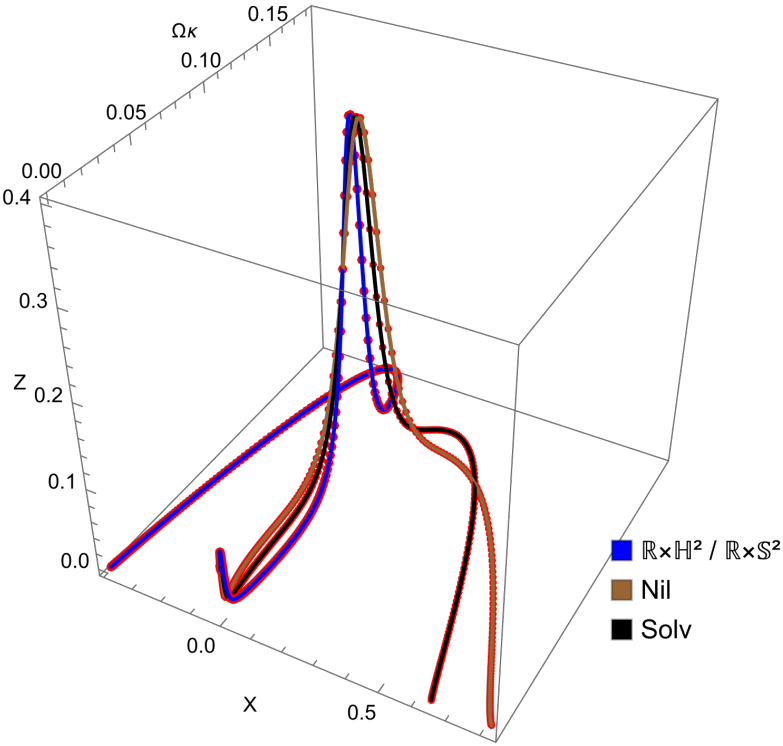}
    \caption{}
    \label{fig:InfPlot1}
\end{subfigure}
\hfill
\begin{subfigure}{.5\textwidth}
    \centering
    \includegraphics[width=0.65\linewidth]{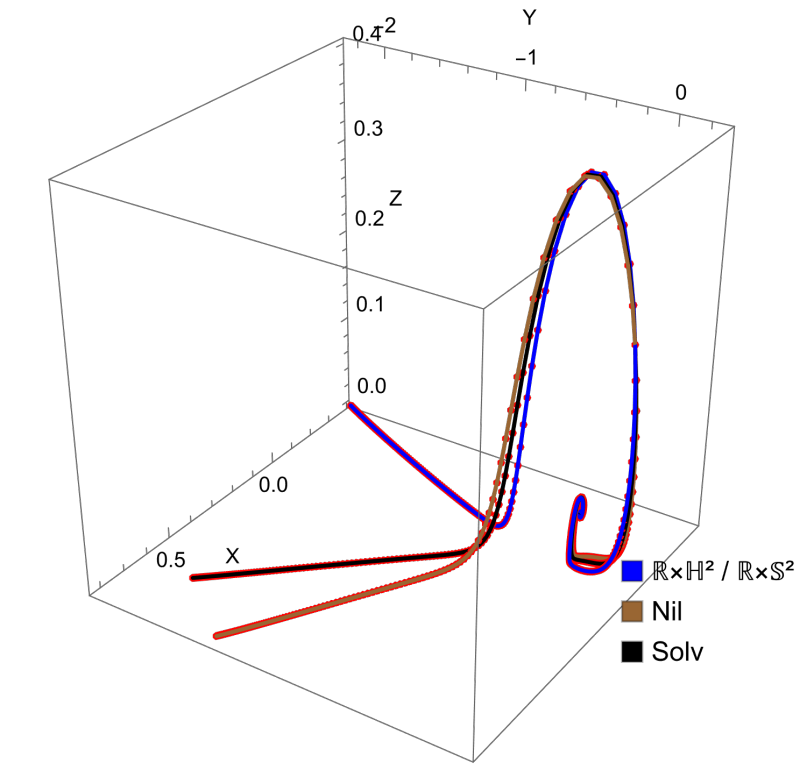}
    \caption{}
    \label{fig:InfPlot2}
\end{subfigure}
\caption{Phase flow plots for (X, Z, $\Omega_\kappa$) (left) and (X, Y, Z) (right) for $\lambda$ = 0.1 and Q = 50 using the initial conditions $X_0$ = 0.01, $Y_0$ = $Z_0$ = 0.25 \& $\Omega_{\kappa,0}$ = 0.1}
\end{figure}
Analysing the phase-space plots from figures \ref{fig:InfPlot1} and \ref{fig:InfPlot2}, we see that the phase space trajectories corresponding to all the considered Thurston spacetimes with an inherent anisotropic curvature converge towards the common attractor point irrespective of the initial conditions. This establishes the existence of a stable anisotropically inflationary fixed point which is eventually approached by the spacetimes governed by Thurston geometries. Therefore, these remarks offer us a solid rationale to conclude the violation of the cosmic no-hair theorem and to declare that inflation with anisotropic hair is cosmologically viable.

%============================================================================================================
\section{Anisotropic phase break} \label{VI}
In this section, we will attempt to track the epoch of anisotropic inflation using phase-space plots and study the evolution of anisotropy as a function of the e-folding number for multiple coupling strengths in Thurston geometries. Following the analyses presented in \cite{byland1998evolution,ito2016mhz}, we will begin with the ansatz defined in equation \eqref{gen metric} and derive its field equations.

We will start with the action defined in equation \eqref{action} of section \ref{inflationary}. Given that for all our anisotropic geometries except $Nil$, the preferred axis lies along z-direction (whereas for $Nil$ it is along x), we then consider the vector field as $A_\mu = (0, A_x(t), 0, 0)$ for \textit{Nil} and $A_\mu = (0, 0, 0, A_z(t))$ for $\mathbb{E}\times \mathbb{H}^2 / S^2$ and \textit{Solv} geometries respectively. \\We have assumed for simplicity that the direction of this vector field does not change in time. This field configuration holds the plane symmetry in the plane perpendicular to the vector.

We consider our anisotropic (\cite{awwad2024large}) Thurston spacetimes
\begin{enumerate}
    \item \textbf{$\mathbb{E} \times \mathbb{H}^2$ and $\mathbb{E} \times S^2$:} 
    \begin{equation} \label{One}
        ds^2 = -dt^2 + e^{2(\alpha + \sigma)}[d\chi^2 + S^2_\kappa(\chi)d\phi^2] + e^{2(\alpha - 2\sigma)}dz^2
    \end{equation}
    \item \textbf{Nil :}
    \begin{equation} \label{Two}
        ds^2 = -dt^2 + e^{2\alpha - 4\sigma}dx^2 + e^{2(\alpha + \sigma)}[(1-\kappa x^2)dy^2 + dz^2 -2x\sqrt{-\kappa}\text{dydz}]
    \end{equation}
    \item \textbf{Solv :}
    \begin{equation} \label{Three}
        ds^2 = -dt^2 + e^{2(\alpha + \sigma)}[e^{2z\sqrt{-\kappa}}dx^2 + e^{-2z\sqrt{-\kappa}}dy^2] + e^{2(\alpha - 2\sigma)}dz^2
    \end{equation}
\end{enumerate}
and write the complete set of evolution equations for them as
\begin{enumerate}
\item \textbf{$\mathbb{E} \times \mathbb{H}^2 / S^2$:} 
    \begin{equation} \label{first}
        \dot{\alpha}^2 - \dot{\sigma}^2 = \frac{1}{3M^2 _{(Pl)}} \left[\frac{\dot{\phi}^2}{2} + V(\phi) + \frac{C e^{-4(\alpha + \sigma)}}{2f^2(\phi)}\right] - \frac{\kappa}{3}e^{-2(\alpha + \sigma)}
    \end{equation}
    \begin{equation}
        \ddot{\alpha} = -3\dot{\alpha}^2 + \frac{1}{M^2 _{(Pl)}} \left[V(\phi) + \frac{Ce^{-4(\alpha + \sigma)}}{6f^2(\phi)}\right] - \frac{2\kappa}{3}e^{-2(\alpha + \sigma)}
    \end{equation}
    \begin{equation}
        \ddot{\sigma} = -3\dot{\alpha}\dot{\sigma} + \frac{C e^{-4(\alpha + \sigma)}}{3 M^2 _{(Pl)} f^2(\phi)} - \frac{\kappa}{3}e^{-2(\alpha + \sigma)}
    \end{equation}
    \begin{equation}
        \ddot{\phi} = -3\dot{\alpha}\dot{\phi} - V'(\phi) + Cf'(\phi)f^{-3}(\phi)e^{-4(\alpha + \sigma)}
    \end{equation}

\vspace{1.5cm}
\item \textbf{Nil :}
    \begin{equation}
        \dot{\alpha}^2 - \dot{\sigma}^2 = \frac{1}{3M^2 _{(Pl)}} \left[\frac{\dot{\phi}^2}{2} + V(\phi) + \frac{C e^{-4(\alpha + \sigma)}}{2f^2(\phi)}\right] - \frac{\kappa}{12}e^{-2\alpha + 4\sigma}
    \end{equation}
    \begin{equation}
        \ddot{\alpha} = -3\dot{\alpha}^2 + \frac{1}{M^2 _{(Pl)}} \left[V(\phi) + \frac{Ce^{-4(\alpha + \sigma)}}{6f^2(\phi)}\right] - \frac{\kappa}{6}e^{-2\alpha + 4\sigma}
    \end{equation}
    \begin{equation}
        \ddot{\sigma} = -3\dot{\alpha}\dot{\sigma} + \frac{C e^{-4(\alpha + \sigma)}}{3 M^2 _{(Pl)} f^2(\phi)} + \frac{\kappa}{6}e^{-2\alpha + 4\sigma}
    \end{equation}
    \begin{equation}
        \ddot{\phi} = -3\dot{\alpha}\dot{\phi} - V'(\phi) + Cf'(\phi)f^{-3}(\phi)e^{-4(\alpha + \sigma)}
    \end{equation}

\vspace{1.5cm}
\item \textbf{Solv :}
    \begin{equation}
        \dot{\alpha}^2 - \dot{\sigma}^2 = \frac{1}{3M^2 _{(Pl)}} \left[\frac{\dot{\phi}^2}{2} + V(\phi) + \frac{C e^{-4(\alpha + \sigma)}}{2f^2(\phi)}\right] - \frac{\kappa}{3}e^{-2\alpha + 4\sigma}
    \end{equation}
    \begin{equation}
        \ddot{\alpha} = -3\dot{\alpha}^2 + \frac{1}{M^2 _{(Pl)}} \left[V(\phi) + \frac{Ce^{-4(\alpha + \sigma)}}{6f^2(\phi)}\right] - \frac{2\kappa}{3}e^{-2\alpha + 4\sigma}
    \end{equation}
    \begin{equation}
        \ddot{\sigma} = -3\dot{\alpha}\dot{\sigma} + \frac{C e^{-4(\alpha + \sigma)}}{3 M^2 _{(Pl)} f^2(\phi)} + \frac{2\kappa}{3}e^{-2\alpha + 4\sigma}
    \end{equation}
    \begin{equation} \label{last}
        \ddot{\phi} = -3\dot{\alpha}\dot{\phi} - V'(\phi) + Cf'(\phi)f^{-3}(\phi)e^{-4(\alpha + \sigma)}
    \end{equation}
\end{enumerate}
where $C>0$ is an arbitrary constant.

From spatial parts of field equations, it is apparent that the rate of anisotropic expansion $\Sigma \equiv \Dot{\sigma}$ depends on the behaviour of the coupling function $f(\phi)$. In the critical case, then, we suppose $f(\phi) \propto \text{e}^{-\text{n}\alpha}$, where n is a parameter characterising coupling strength (see appendix \ref{ap2}).

\begin{figure} [h!]
    \begin{subfigure}{0.5\textwidth}           
    \centering
    \includegraphics[width=0.7\linewidth]{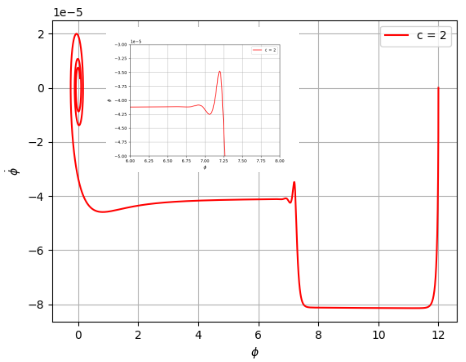}
    \caption{$\mathbb{E} \times \mathbb{H}^2$}
    \label{V2-both-rh2-1}
    \end{subfigure}
\hfill
    \begin{subfigure}{0.5\textwidth}           
    \centering
    \includegraphics[width=0.7\linewidth]{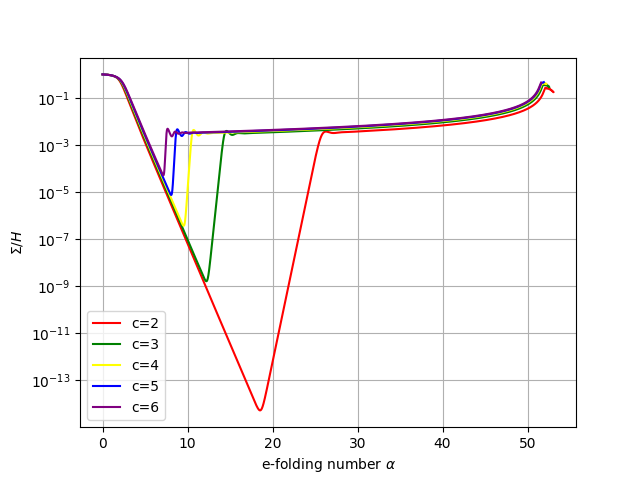}
    \caption{$\mathbb{E} \times \mathbb{H}^2$}
    \label{V2-both-rh2-2}
    \end{subfigure}
\hfill
    \begin{subfigure}{0.5\textwidth}        
    \centering
    \includegraphics[width=0.7\linewidth]{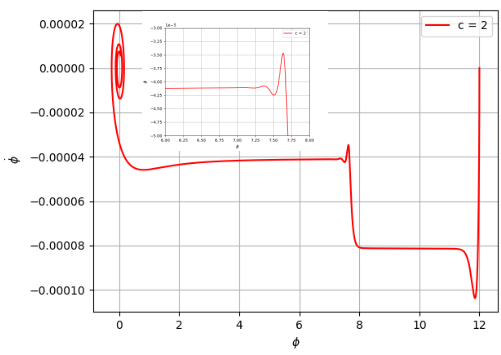}
    \caption{$\mathbb{E} \times S^2$}
    \label{V2-both-rs2-1}
    \end{subfigure}
\hfill
    \begin{subfigure}{0.5\textwidth}        
    \centering
    \includegraphics[width=0.7\linewidth]{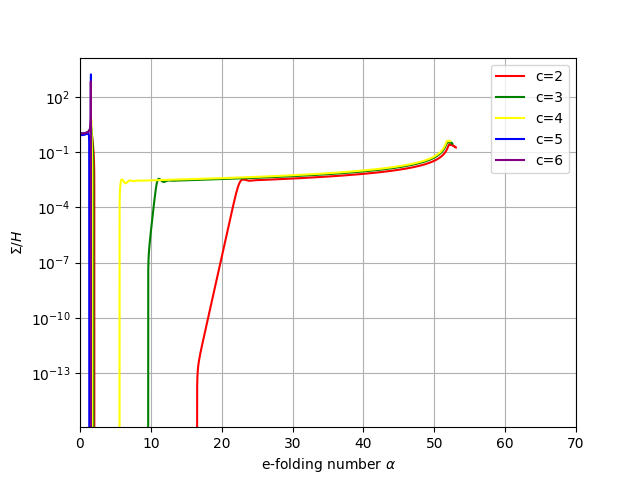}
    \caption{$\mathbb{E} \times S^2$}
    \label{V2-both-rs2-2}
    \end{subfigure}
\hfill
    \begin{subfigure}{0.5\textwidth}
    \centering
    \includegraphics[width=0.7\linewidth]{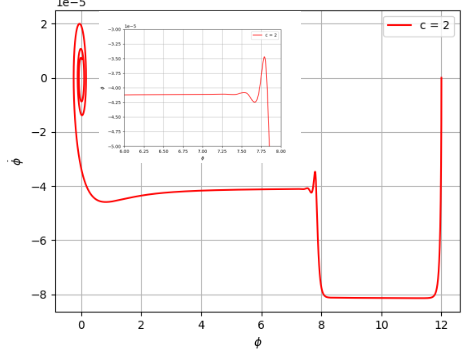}
    \caption{Nil \& Solv}
    \label{V2-both-nil-1}
    \end{subfigure}
\hfill
    \begin{subfigure}{0.5\textwidth}
    \centering
    \includegraphics[width=0.7\linewidth]{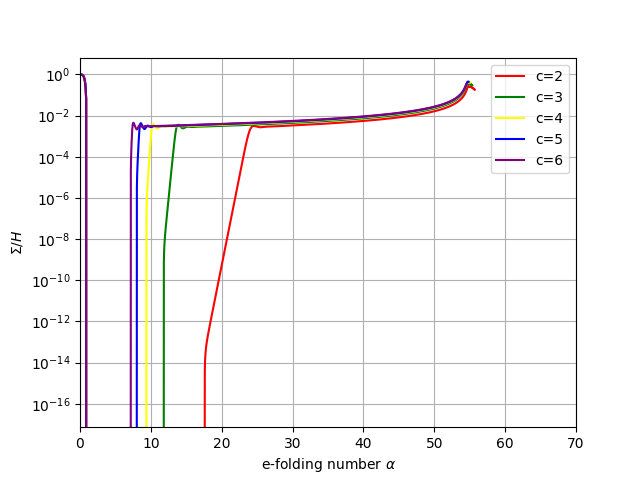}
    \caption{Nil \& Solv}
    \label{V2-both-nil-2}
    \end{subfigure}
\caption{Phase space (left) and anisotropy evolution (right) plots for potential $V(\phi) = \frac{1}{2}m^2 \phi^2$}
\label{powerlaw}
\end{figure}

\begin{figure} [h!]
    \begin{subfigure}{0.5\textwidth}           
    \centering
    \includegraphics[width=0.7\linewidth]{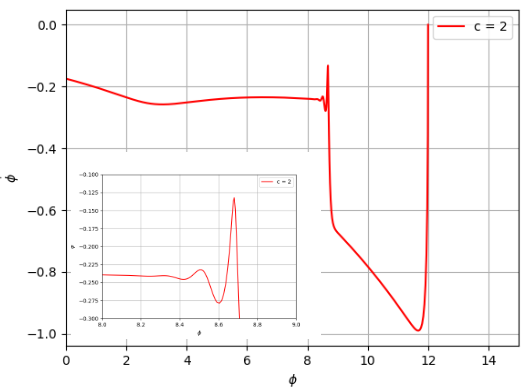}
    \caption{All geometries}
    \label{V1-both-rh2-1}
    \end{subfigure}
\hfill
    \begin{subfigure}{0.5\textwidth}           
    \centering
    \includegraphics[width=0.7\linewidth]{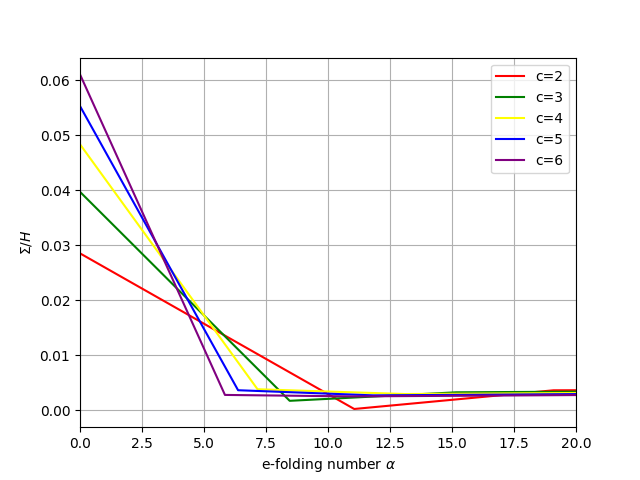}
    \caption{All geometries}
    \label{V1-both-rh2-2}
    \end{subfigure}
\caption{Fig. \ref{powerlaw} for potential $V(\phi) = V_0 e^\frac{\lambda \phi}{M_{(Pl)}}$}
\label{exponential}
\end{figure}

Substituting values of equations \eqref{V-form} \& \eqref{f-form} (see appendix \ref{ap2}) (using equation \eqref{A}) into equations \eqref{first} - \eqref{last} and solving them numerically, we obtain for two choices of scalar field potentials - exponential (\cite{Hervik_2011})
\begin{equation}
    V(\phi) = V_0 e^\frac{\lambda \phi}{M_{(Pl)}}
\end{equation}
and power law (\cite{Watanabe_2009, ito2016mhz, soda2012statistical})
\begin{equation}
    V(\phi) = \frac{1}{2}m^2\phi^2
\end{equation}
the \textit{phase space} and \textit{anisotropy evolution} plots, shown in figs. \ref{powerlaw} - \ref{exponential} using our redefined coupling parameter (lowercase) c = n/2. The initial values used are:
\begin{multicols}{2}
\begin{itemize}
    \item $V_0$ = $M_{(Pl)}$ = 1
    \item $\lambda$ = 0.1
    \item m = $10^{-4}M_{(Pl)}$
    \item n = 4
    \item $|\kappa| = 10^{-3}$
    \item $CC^{-2} _1 = e^{98}$
    \item $\dot{\alpha}$(0) $\to$ From constraint equation
    \item $\alpha$(0) = $\sigma$(0) = $\dot{\phi}$(0) = 0
    \item $\dot{\sigma}$(0) = 0.1
    \item $\phi$(0) = $12M_{(Pl)}$
    \item t = 0 to $4 \times 10^5$
    \item time points = $10^5$
    \item (rtol, atol) = ($10^{-9}$, $10^{-12}$)
\end{itemize}
\end{multicols}
In the phase space plots for the power law potential (figures \ref{V2-both-rh2-1}, \ref{V2-both-rs2-1} \& \ref{V2-both-nil-1}), we observe that the open geometries have a \textit{smoother} phase of isotropic inflation ($\phi \simeq$ 12 to 8) with Nil \& Solv behaving \textit{exactly} in an identical way. On the other hand, closed geometries such as $\mathbb{E} \times S^2$ exhibit a small \textit{kink} at the onset of isotropic inflation, lasting for an extremely brief period, after which smoothness is restored, similar to their hyperbolic counterparts. All the geometries converge to an attractor at the end of anisotropic inflation ($\phi$ = 0).

In the anisotropy evolution plots for the power law potential (figures \ref{V2-both-rh2-2}, \ref{V2-both-rs2-2} \& \ref{V2-both-nil-2}), we observe that all the geometries converge to an anisotropic attractor point after about 50 e-folds. $\mathbb{E} \times \mathbb{H}^2$ behaves very uniquely from other Thurston geometries in the sense that its coupling strength parameter \textit{c} \textit{counter-affects} the brief dip seen in anisotropy starting from about 10 e-folds and continuing till around 20 e-folds, i.e., the \textit{higher} the coupling parameter in $\mathbb{E} \times \mathbb{H}^2$, the \textit{shorter} the duration of anisotropic dip observed in the spacetimes governed by this geometry (see figure \ref{V2-both-rh2-2}). The situation, however, is different for other geometries, as the \textit{turning point} for such an anisotropic dip is \textit{unknown} and hence the anisotropies (corresponding to different coupling strengths but especially for c = 2, 3, 4) \textit{spontaneously} appear into the picture at specific e-fold numbers and actually extend well upto an enormous negative integer if one is willing to extrapolate the y labels in figures \ref{V2-both-rs2-2} and \ref{V2-both-nil-2} well upto minus infinity. In figure \ref{V2-both-nil-2}, even the anisotropic curves corresponding to higher coupling strength parameters are visible as compared to those of $\mathbb{E} \times S^2$, but behave similarly as for the spontaneous origin of anisotropy at select e-fold numbers. The rise amount, however, in all the geometries is extremely small, and the only physical observation that matters is the impulsive origins of anisotropies corresponding to various e-fold numbers in all the anisotropic Thurston geometries that converge eventually to a single stable attractor. 

We also plotted the same figures for the exponential potential (equation \eqref{expoten}) in Figure \ref{exponential}, and the results were quite comprehensive. Remarkably, all the geometries are now mutually indistinguishable. The phase space plots \ref{V1-both-rh2-1} are slightly different from those we saw in the power law potential since the isotropic phase ($\phi \simeq$ 12 to 9) is not particularly stable, but transitions steadily to an anisotropic one after which it attains stability all its way up to $\phi$ = 0. We urge the reader to refer to the ref. \cite{soda2012statistical}.

Finally, the anisotropy evolution plots \ref{V1-both-rh2-2} are also identical for all the anisotropic Thurston geometries in this potential, but behave in a different way as compared to their power law counterparts. The anisotropy indeed converges to a non-zero stable attractor point only after 20 e-folds, but there exist nonzero initial anisotropies in all the geometries for all the coupling strength parameters. Specifically, this initial anisotropy was found to be (approximately) directly proportional to the coupling strength parameters themselves. The anisotropies decay linearly well up to about 8 e-folds, on average, to a small but non-zero value, after which they converge to their attractor point.


%============================================================================================================
\newpage
\section{Results \& Analysis} \label{VII}
The inflationary modelling of the Thurston spacetimes led to an analysis of the existence and dynamics of an anisotropic inflationary epoch in the universe. The formulation of dynamic phase space systems for the Thurston spacetimes, derived from their field equations, and subsequent stability evaluations demonstrate that each of these geometries supports the stable existence of an anisotropic inflation. The phase flow analysis from figures \ref{fig:InfPlot1} and \ref{fig:InfPlot2} in section \ref{Phase flow analysis} leaves a clear indication of these remarks, which will also be interpreted from the following section. 

From our observations in Section \ref{VI}, corresponding to the results shown in Figures \ref{powerlaw} - \ref{exponential}, we analyse a few important aspects. Foremost, the plots are \textit{potential-dependent}: a power-law potential, a specific case of an exponential potential, allows one to differentiate among geometries, whereas the more general exponential potential hard-codes them. Further, in the power law potential, only the Thurston geometry having a positive cylindrical curvature ($\mathbb{E} \times S^2$) has a sharp \textit{kink} in the beginning of phase space plot which is purely a characteristic of its inherent spatial curvature, since all the rest of the geometries carrying a negative curvature behave roughly the same. Talking of this, one can even distinguish among those with negative spatial curvatures by observing the \textit{field lag} between the two epochs of cosmic inflation. Specifically, in Cartesian open geometries (Nil \& Solv), the transition is almost instantaneous while in cylindrical open geometry ($\mathbb{E} \times \mathbb{H}^2$), there is a noticeable lag of about $\phi \simeq$ 1 units, suggesting that not only the spatical curvature but also the \textit{inherent geometry} of the spacetime itself plays a vital role in determining the physics of early universe. 

It is interesting to see that this latter case of an open cylindrical geometry ($\mathbb{E} \times \mathbb{H}^2$) is the only one with a smooth anisotropy transition with e-folding numbers among all the anisotropic Thurston geometries. Indeed, a stable sttractor exists in all our geometries, but the origin is \textit{sudden} and \textit{spontaneous} in all the other geometries except $\mathbb{E} \times \mathbb{H}^2$. Also, the weaker the coupling (value of \textit{c}), the later the anisotropy starts to rise with the e-fold number. This directly verifies the presence of an anisotropy-causing axisymmetric vector field.

The exponential potential, being much more rapidly varying in $\phi$ than the power-law potential, does not directly allow one to distinguish among geometries. It is a characteristic of this potential that a \textit{stable} isotropic inflationary phase does \textit{not} exist; $\phi$ steadily increases (through what in a power law potential was a \textit{stable} isotropic epoch) and settles at a steady anisotropic one, from where it continues to its stable attractor point.

The exponential potential also smoothes out anisotropies to a greater extent than the power law, as the stable attractor point has about 10 times smaller anisotropies than the power law one. We also see that for this potential, the weaker the coupling strength, the lower the initial anisotropy is, thus re-verifying the presence of our axisymmetric vector field. Further, the exponential potential has also been observed to force the anisotropies to decay \textit{strongly linearly} up to about 8 e-folds, from where they traverse towards their stable attractor point.

All such observations clearly point to a residual \textit{anisotropic stable attractor point} post anisotropic inflationary era. Thus. The cosmic no-hair theorem is clearly \textit{violated} for the case of \textit{Thurston spacetimes} for both power law and exponential potentials.

%============================================================================================================
\section{Conclusion} \label{VIII}
This paper focuses on the inflationary epoch of the Universe by modelling and analysing its underlying geometry within the framework of Thurston spacetimes. In doing so, it examines the validity of the cosmic no-hair theorem for this class of geometries and provides insight into the coupling of the resulting vector field arising from these spacetimes with an otherwise standard inflationary scenario. 

After developing inflationary models for the considered set of Thurston geometries, we formulated them as dynamical systems for analysis in phase space. We employed the linear stability approach to demonstrate the existence of a stable anisotropic fixed point, unique to the entire set of Thurston spacetimes, given by the point A.

The phase space trajectories for each of the considered Thurston spacetimes with arbitrary initial conditions converge towards the common fixed point A, indicating that the universe will get close to this attractor after a few e-folds of inflation. Therefore, the stable and anisotropically inflationary fixed point imbibe a major fraction of e-folds of the inflationary epoch of the universe.

Our analysis has become more specific and general through understanding the existence of \textit{two} slow-roll phases, corresponding to \textit{isotropic} epochs, followed by \textit{anisotropic} epochs. The evolution of this anisotropic phase has been traced throughout a sufficient number of e-foldings of inflation. Hence, we address the existence of anisotropic inflation in the universe and present a \textit{counter example} to the theoretical cosmic no-hair theorem. 

Remarkably, in \cite{smith2025cosmological}, an upper bound on $\Omega_\kappa$ was constrained on Thurston geometries after a sufficient evolution of the scale factor, thereby presenting a proof of the violation of the cosmic no-hair theorem similar to our case. Other works, such as \cite{Watanabe_2009, soda2012statistical, ito2016mhz, maleknejad2012revisiting} related to ours, also establish a counterexample to the cosmic no-hair conjecture, along with the dependence of late-time evolution and not the inflationary era itself on the potential for the inflation field. More importantly, the works \cite{ito2016mhz, franco2017tensor, hertzberg2024constraints} explore the production of copious primordial gravitational waves, tensor perturbations \& the time evolution of anisotropic shear, which can be very well translated for the unexplored case of Thurston geometries so far, which we plan to carry out in future.

%============================================================================================================
\newpage
\begin{appendices}
\section{Critical points}
\subsection{ $\mathbb{E}\times \mathbb{H}^2$ and $\mathbb{E} \times S^2$ geometry}\label{app:RH2 app}
\vspace{10 pt}
\textbf{Critical point A:} Considered under the positive energy constraint of the vector field $\Omega_A >0$, implying $Z^2 >0$. We infer that the solution A exists only in the region where $\lambda^2 + 2Q\lambda > 4$. As far as $\lambda<<1$, this shows $Q>>1$ and also the potential energy of the scalar field (due to the Y variable) dominates this fixed point. Given in table \ref{tab:InflationRH2FP} is the approximation of the actual fix point to the lowest order in $\frac{\lambda}{Q}$ and $Q^{-2}$ when $\lambda<<1$ and $Q>>1$. The exact position of A is:
\begin{subequations}
    \begin{gather}
        X= \frac{2 (-4 + 2Q\lambda + \lambda^2)}{8 + 12Q^2 + 8Q\lambda + \lambda^2}\\
Y= -\frac{12 (2Q + \lambda)}{8 + 12Q^2 + 8Q\lambda + \lambda^2}\\
Z^2 = \frac{18 \left(-32 + 24 Q^3 \lambda + 12 \lambda^2 + 2 Q \lambda^3 - \lambda^4 + 4 Q^2 (-12 + 5 \lambda^2)\right)}{(8 + 12 Q^2 + 8 Q \lambda + \lambda^2)^2}\\
\Omega_\kappa = 0
\end{gather}
\end{subequations}
The eigenvalues of the point when expanded in terms of $\frac{\lambda}{Q}$ and $Q^{-2}$ and truncated at zero order will give: 
\begin{equation} 
\left\{ -2, -3, \frac{-3}{2} - \frac{i}{2} \sqrt{3(-19+8Q\lambda)}, \frac{-3}{2} + \frac{i}{2} \sqrt{3(-19+8Q\lambda)} \right\}
\end{equation}
The negative eigenvalues -2 and -3, and the negative real parts of the remaining complex eigenvalues, conclude A as a \textit{stable} fixed point.\\

\textbf{Critical point B:} An inflationary fixed point for the spacetime considered consisting only of a scalar field predominated by potential energy (Y variable). The eigenvalues of B truncated at $\mathcal{O}(\lambda)$ would turn out as:
\begin{equation}
    \{-3,-3,-2,-2+Q\lambda\}
\end{equation}
The stability analysis implies that for $Q\lambda >2$, B is a \textit{saddle}, while $Q\lambda<2$ would give all negative eigenvalues and so, B will be a \textit{stable} point.\\

\textbf{Critical point C:} Exists only in the parameter region $\frac{\lambda}{Q}<2$ due to the positive energy condition of the vector field giving $Z^2 >0$. The eigenvalues of the fixed point truncated at first order in $\lambda$ are:
\begin{equation}
    \{3, -6, -3 + 6 Q \lambda, -6 Q \lambda\}
\end{equation}
and therefore, C is a \textit{saddle} point.\\

\textbf{Critical point D:} An inflationary fixed point, containing only the scalar field $(Z=0)$. The eigenvalues:
\begin{equation}
    \{3 Q \lambda, -3, -6, 3\},
\end{equation}
imply D to be a \textit{saddle} point.\\

\textbf{Critical point E:} The trajectory of fixed points satisfying the equation  $6X^2 + Y^2 =1$ and lies on the boundary of the state-space of our system, as the constraint equation \eqref{constraintRinf} equals 1 (i.e., $V(\phi)=0)$. For the eigenvalues:
\begin{equation}
    \{0, -2 (-2 + X), 1 - 2X \mp \sqrt{6} Q \sqrt{1 - X^2}, 
 6 \pm \sqrt{6} \sqrt{1 - X^2} \lambda\}
\end{equation}
We have 0 as an eigenvalue, and we refer to these critical points as \textit{non-hyperbolic} points. In these cases, we look for the presence of any positive eigenvalues that can make the unstable space non-empty and hence conclude the stability of the non-hyperbolic point to be unstable.
Therefore, here, E is an \textit{unstable} fixed point.\\

\textbf{Critical point F:} Also a boundary point that contains contributions only from the shear and curvature variables, which means it has \textit{no} fluids. The q value of $\frac{1}{2}$ makes it a decelerating fixed point, and the following eigenvalues imply F to be a \textit{saddle}:
\begin{equation}
    \left\{3, -\frac{3}{2}, -\frac{3}{2}, -\frac{3}{2}\right\}
\end{equation}\\

\textbf{Critical point G:} A decelerating, \textit{unstable} boundary point under the condition $Q>>1$, with eigenvalues: 
\begin{equation}
    \left\{0, 0, 2+
\frac{1}{6Q^2}
6 -\frac{\lambda}{Q}  \right\}
\end{equation}\\

\textbf{Critical point H:} A decelerating fixed point that lies on the boundary, which is \textit{unstable} for $Q>>1$ as concluded from the eigenvalues:
\begin{equation}
     \left\{ 0, 0,6- \frac{3}{2Q^2}, 6+\frac{3\lambda}{Q} \right\}
\end{equation}

\subsection{Nil geometry}\label{app:Nil app}
\vspace{10 pt}
\textbf{Critical point A:} Similar to that of the point A in  $\mathbb{E}\times \mathbb{H}^2$ and $\mathbb{E} \times S^2$, and therefore A exists only in the region where $\lambda^2 + 2Q\lambda > 4$ and  since $\lambda<<1$, it implies $Q>>1$. The exact position of A is:
\begin{subequations}
    \begin{gather}
        X= \frac{2 (-4 + 2Q\lambda + \lambda^2)}{8 + 12Q^2 + 8Q\lambda + \lambda^2}\\
Y= -\frac{12 (2Q + \lambda)}{8 + 12Q^2 + 8Q\lambda + \lambda^2}\\
Z^2 = \frac{18 \left(-32 + 24 Q^3 \lambda + 12 \lambda^2 + 2 Q \lambda^3 - \lambda^4 + 4 Q^2 (-12 + 5 \lambda^2)\right)}{(8 + 12 Q^2 + 8 Q \lambda + \lambda^2)^2}\\
\Omega_\kappa = 0
\end{gather}
\end{subequations}
The eigenvalues of the point are approximated as: 
\begin{equation} 
\left\{ -2, -3, \frac{-3}{2} - \frac{i}{2} \sqrt{3(-19+8Q\lambda)}, \frac{-3}{2} + \frac{i}{2} \sqrt{3(-19+8Q\lambda)} \right\}
\end{equation}
such that A is a \textit{stable} and inflationary fixed point.\\

\textbf{Critical point B:} Only consists of a scalar field primarily influenced by its potential energy. The eigenvalues of B truncated at $\mathcal{O}(\lambda)$
\begin{equation}
    \{-3,-3,-2,-2+Q\lambda\}
\end{equation}
imply that B will be a \textit{stable}, inflationary point for $Q\lambda<2$ and for $Q\lambda >2$ it is a \textit{saddle}.\\

\textbf{Critical point C:} It is the approximation of the exact position of the point in table \ref{tab:InflationNilFP} at:
\begin{subequations}
    \begin{gather}
        X= \left(\frac{2Q - \lambda}{4(Q + \lambda)}\right)\\
        Y = -\frac{3}{Q + \lambda}\\
        Z^2=\frac{9(12Q^2\lambda+8Q(-4+3\lambda^2)+\lambda(-32+9\lambda^2))}{8(Q + \lambda)^2(4Q+3\lambda)}\\
        \Omega_\kappa=-\frac{3(8 + 4Q^2 - 4Q\lambda - 3\lambda^2)}{8(Q + \lambda)(4Q+3\lambda)} \end{gather}
\end{subequations}\\
The point satisfies the constraint equation \eqref{constraintNilinf} under the condition $Q >> 1$. The eigenvalues of the fixed point would therefore be approximated as:
\begin{equation}
    \left\{-3, -\frac{45}{8}, \frac{15}{8}, \frac{27}{8}\right\}
\end{equation}
implying that C is a \textit{saddle}. Also, the point is that inflation is inflationary under the given conditions.\\

\textbf{Critical point D:} Lies on the boundary of the state space of the \textit{Nil} system as the constraint equation \eqref{constraintNilinf} equals 1. The eigenvalues:
\begin{equation}
    \{0, 4 (1 + X), 1 - 2X \mp \sqrt{6} Q \sqrt{1 - X^2}
 6 \pm \sqrt{6} \sqrt{1 - X^2} \lambda\}
\end{equation}
imply that D is an \textit{unstable} point and since $q = 2 > 0$, it will be a \textit{decelerating} fixed point.\\

\textbf{Critical point E:} It is an \textit{unstable}, decelerating boundary point for $Q >> 1$ as concluded from the given eigenvalues:
\begin{equation}
     \left\{ 0, 0, 8-\frac{1}{3Q^2}, 6-\frac{\lambda}{Q} \right\}
\end{equation}\\

\textbf{Critical point F:} It is also an \textit{unstable}, decelerating boundary point for $Q >> 1$ with eigenvalues:
\begin{equation}
     \left\{ 0, 0, \frac{3}{Q^2}, 6+\frac{3\lambda}{Q} \right\}
\end{equation}\\

\textbf{Critical point G:} It will also be an \textit{unstable}, decelerating boundary point for $Q >> 1$ with eigenvalues:
\begin{equation}
     \left\{ 6, 3, 0, 0 \right\}
\end{equation}\\

\textbf{Critical point H:} It will be a \textit{stable}, inflationary boundary point with eigenvalues:
\begin{equation}
     \left\{ -18, -9, -6, -6 \right\}
\end{equation}\\

\textbf{Critical point I:} It is a \textit{saddle}, decelerating boundary point for $Q >> 1$ with eigenvalues:
\begin{equation}
     \left\{ 6-\frac{4}{Q^2}+\frac{3\lambda}{Q}, \frac{-3}{Q^2}, \frac{-1}{Q^2}-\frac{2\sqrt{3}i}{Q},\frac{-1}{Q^2}+\frac{2\sqrt{3}i}{Q} \right\}
\end{equation}\\

\subsection{Solv geometry}\label{app:Solv app}
\vspace{10 pt}
\textbf{Critical point A:} Given in table \ref{tab:InflationSolvFP}, this is the same as point A in the previous sections truncated at the lowest order in $\lambda$ and $Q^{-2}$ and exists only in the region where $\lambda^2 + 2Q\lambda > 4$, implying $Q >> 1$, and the domination of the potential energy of inflation.\\
The eigenvalues of A are approximated as: 
\begin{equation} 
\left\{ -2, -3, \frac{-3}{2} - \frac{i}{2} \sqrt{3(-19+8Q\lambda)}, \frac{-3}{2} + \frac{i}{2} \sqrt{3(-19+8Q\lambda)} \right\}
\end{equation}
such that A would be a \textit{stable} point. Under the conditions $\lambda << 1$ and $Q >> 1$, the deceleration parameter, $q < 0$ indicates A to be an inflationary fixed point.\\

\textbf{Critical point B:} Consists only of a scalar field due to the Y variable contribution in the state space. The eigenvalues of B truncated at the lowest order in $\lambda$ would be:
\begin{equation}
    \{-3,-3,-2,-2+Q\lambda\}
\end{equation}
implying B to be a \textit{stable}, inflationary point for $Q\lambda < 2$ and \textit{saddle} for $Q\lambda > 2$.\\

\textbf{Critical point C:} In table \ref{tab:InflationSolvFP}, this is the approximation of its exact position at:
\begin{subequations}
    \begin{gather}
        X= \left(\frac{2Q - \lambda}{4(Q + \lambda)}\right)\\
        Y = -\frac{3}{Q + \lambda}\\
        Z^2=\frac{9(-4 + 2Q\lambda + \lambda^2)}{4(Q + \lambda)^2}\\
        \Omega_\kappa=-\frac{3(8 + 4Q^2 - 4Q\lambda - 3\lambda^2)}{16(Q + \lambda)^2} \end{gather}
\end{subequations}\\
The point also considers the positive energy condition of the vector field $Z^2 > 0$, which gives $\lambda^2 + 2Q\lambda > 4$, and therefore the existence would be valid for $Q >> 1$. The eigenvalues of the fixed point would therefore be approximated as:
\begin{equation}
    \left\{-6, -\frac{3}{2}(1+\sqrt{5}), \frac{3}{2}(-1+\sqrt{5}), 3\right\}
\end{equation}
implying that C is a \textit{saddle}. Also, the point is that inflation is inflationary under the given conditions.\\

\textbf{Critical point D:} It is a \textit{saddle} inflationary fixed point with the following eigenvalues:
\begin{equation}
    \left\{-3+\frac{3Q\lambda}{2}, -3, -\frac{3}{2}(1+\sqrt{5}), \frac{3}{2}(-1+\sqrt{5})\right\}
\end{equation}\\

\textbf{Critical point E:} This is a boundary point on the state space of the \textit{Solv} system. The eigenvalues correspond to:
\begin{equation}
    \{0, 4 (1+ X), 1 - 2X \mp \sqrt{6} Q \sqrt{1 - X^2}
 6 \pm \sqrt{6} \sqrt{1 - X^2} \lambda\}
\end{equation}
such that E is an \textit{unstable} fixed point and $q = 2 > 0$, so E will be a \textit{decelerating} fixed point.\\

\textbf{Critical point F:} This is a \textit{decelerating} fixed point lying on the boundary that contains only the shear part. The following eigenvalues imply F to be an \textit{unstable} point:
\begin{equation}
   \left\{ 6,3,0,0 \right\}
\end{equation}\\

\textbf{Critical point G:} This is a decelerating fixed point on the boundary of the state space, with the following eigenvalues (truncated at the lowest order in $\frac{\lambda}{Q}$ and $Q^{-2}$) show G to be a \textit{saddle} for $Q >> 1$:
\begin{equation}
    \left\{ -\frac{3}{Q^2}, \quad 6 - \frac{6}{Q^2} + \frac{3\lambda}{Q} \quad 
-\frac{9}{2Q^2} - \frac{3i\sqrt{2}}{Q} \quad 
-\frac{9}{2Q^2} + \frac{3i\sqrt{2}}{Q} \right\}
\end{equation}\\

\textbf{Critical point H:} This is an \textit{unstable, decelerating} boundary point for $Q >> 1$ as observed from the eigenvalues:
\begin{equation}
     \left\{ 0, 0, 8 - \frac{1}{3Q^2}, 6-\frac{\lambda}{Q} \right\}
\end{equation}\\

\textbf{Critical point I:} It is also an \textit{unstable, decelerating} fixed point for $Q >> 1$ lying on the boundary of the state space with eigenvalues:
\begin{equation}
     \left\{ 0, 0, \frac{3}{Q^2}, 6+\frac{3\lambda}{Q} \right\}
\end{equation}\\

%============================================================================================================
\section{Coupling function \& anisotropic condition}\label{ap2}
\vspace{20 pt}
From \cite{ito2016mhz}, the condition for the occurrence of anisotropic inflation reads
\begin{equation} \label{condition-1}
    M^2 _{\text(Pl)}\frac{f'(\phi)}{f(\phi)}\frac{V'(\phi)}{V(\phi)} > 2
\end{equation}
Then, considering one Thurston geometry at a time, assuming slow-roll and ignoring the backreaction of the field in the field equations, we have from the reduced Einstein \& scalar field equations,
\begin{equation}
    3\Dot{\alpha}^2 \simeq \frac{V(\phi)}{M^2 _\text{Pl}}
\end{equation}
and
\begin{equation}
    3\Dot{\alpha\phi} \simeq -V'(\phi)
\end{equation}
so that, upon dividing, we have
\begin{equation}
    \frac{d\alpha}{d\phi} = -\frac{1}{M^2 _\text{Pl}}\frac{V(\phi)}{V'(\phi)}
\end{equation}
Integrating, we get
\begin{equation}
    \boxed{\alpha(\phi) = -\frac{1}{M^2 _{\text(Pl)}} \int \frac{V(\phi)}{V'(\phi)}d\phi + C_0}
\end{equation}
where $C_0$ is an arbitrary constant of integration. For various geometries, we then have
\begin{equation} \label{131}
\begin{split}
    f(\phi) &= e^{-n\alpha}\\
    &= C_1 e^{\frac{n}{M^2 _{\text{Pl}}} \int \frac{V(\phi)}{V'(\phi)}d\phi}
\end{split}
\end{equation}
where $C_1 = e^{C_0}$.\\
Considering chaotic inflation with the potential
\begin{equation} \label{V-form}
    V(\phi) = \frac{1}{2}\text{m}^2\phi^2
\end{equation}
where m ($\approx 10^{-5}$) denotes the mass of the inflation field, we solve for $f(\phi$) in equation \eqref{131} to obtain
\begin{equation} \label{f-form}
\begin{split}
    f(\phi) &= C_1 e^{\frac{n}{M^2 _{(Pl)}} \int \frac{\phi}{2} d\phi}\\
    &= C_1 e^{\frac{n}{M^2 _{(Pl)}} \frac{\phi^2}{4}}\\
    &= \boxed{C_1 e^{n\phi^2/4M^2 _{(Pl)}}}
\end{split}
\end{equation}
Thus substituting values of V($\phi$) and f($\phi$) from equations \eqref{V-form} and \eqref{f-form} respectively into equation \eqref{condition-1}, we obtain for our various anisotropic geometries,
\begin{equation}
    \boxed{n > 2}
\end{equation}
\end{appendices}

\bibliography{main}

@article{aghanim2020planck,
  title={Planck 2018 results-I. Overview and the cosmological legacy of Planck},
  author={Aghanim, Nabila and Akrami, Yashar and Arroja, Frederico and Ashdown, Mark and Aumont, J and Baccigalupi, Carlo and Ballardini, M and Banday, Anthony J and Barreiro, RB and Bartolo, Nicola and others},
  journal={Astronomy \& Astrophysics},
  volume={641},
  pages={A1},
  year={2020},
  publisher={EDP sciences}
}

@article{awwad2024large,
  title={Large-scale geometry of the Universe},
  author={Awwad, Yassir and Prokopec, Tomislav},
  journal={Journal of Cosmology and Astroparticle Physics},
  volume={2024},
  number={01},
  pages={010},
  year={2024},
  publisher={IOP Publishing}
}

@article{Hervik_2011,
   title={Inflation with stable anisotropic hair: is it cosmologically viable?},
   volume={2011},
   ISSN={1029-8479},
   url={http://dx.doi.org/10.1007/JHEP11(2011)146},
   DOI={10.1007/jhep11(2011)146},
   number={11},
   journal={Journal of High Energy Physics},
   publisher={Springer Science and Business Media LLC},
   author={Hervik, Sigbjørn and Mota, David F. and Thorsrud, Mikjel},
   year={2011},
   month=nov }

@article{Watanabe_2009,
   title={Inflationary Universe with Anisotropic Hair},
   volume={102},
   ISSN={1079-7114},
   url={http://dx.doi.org/10.1103/PhysRevLett.102.191302},
   DOI={10.1103/physrevlett.102.191302},
   number={19},
   journal={Physical Review Letters},
   publisher={American Physical Society (APS)},
   author={Watanabe, Masa-aki and Kanno, Sugumi and Soda, Jiro},
   year={2009},
   month=may }

@article{watanabe2011imprints,
  title={Imprints of the anisotropic inflation on the cosmic microwave background},
  author={Watanabe, Masa-aki and Kanno, Sugumi and Soda, Jiro},
  journal={Monthly Notices of the Royal Astronomical Society: Letters},
  volume={412},
  number={1},
  pages={L83--L87},
  year={2011},
  publisher={Blackwell Publishing Ltd Oxford, UK}
}

@incollection{ellis1999cosmological,
  title={Cosmological models: Cargese lectures 1998},
  author={Ellis, George FR and Van Elst, Henk},
  booktitle={Theoretical and Observational Cosmology},
  pages={1--116},
  year={1999},
  publisher={Springer}
}

@article{thurston1982three,
  author    = {William P. Thurston},
  title     = {Three-dimensional manifolds, Kleinian groups and hyperbolic geometry},
  journal   = {Bulletin of the American Mathematical Society (New Series)},
  volume    = {6},
  number    = {3},
  pages     = {357--381},
  year      = {1982},
  month     = {May},
  doi       = {10.1090/S0273-0979-1982-15003-0},
  publisher = {American Mathematical Society}
}

@article{perelman2003finite,
  title={Finite extinction time for the solutions to the Ricci flow on certain three-manifolds},
  author={Perelman, Grisha},
  journal={arXiv preprint math/0307245},
  year={2003}
}

@article{perelman2002entropy,
  title={The entropy formula for the Ricci flow and its geometric applications},
  author={Perelman, Grisha},
  journal={arXiv preprint math/0211159},
  year={2002}
}

@article{perelman2003ricci,
  title={Ricci flow with surgery on three-manifolds},
  author={Perelman, Grisha},
  journal={arXiv preprint math/0303109},
  year={2003}
}

@techreport{kitada199207579cosmic,
  title={Cosmic no hair theorem in exponential and power law inflation: Extended Wald’s theorem},
  author={Kitada, Y and Maeda, K-i},
  institution={UTAP-137-92, http://ccdb5fs. kek. jp/cgi-bin/img/allpdf}
}

@article{Bahamonde_2018,
   title={Dynamical systems applied to cosmology: Dark energy and modified gravity},
   volume={775–777},
   ISSN={0370-1573},
   url={http://dx.doi.org/10.1016/j.physrep.2018.09.001},
   DOI={10.1016/j.physrep.2018.09.001},
   journal={Physics Reports},
   publisher={Elsevier BV},
   author={Bahamonde, Sebastian and Böhmer, Christian G. and Carloni, Sante and Copeland, Edmund J. and Fang, Wei and Tamanini, Nicola},
   year={2018},
   month=nov, pages={1–122} }

@article{byland1998evolution,
  title={Evolution of the Bianchi type I, Bianchi type III, and the Kantowski-Sachs universe: Isotropization and inflation},
  author={Byland, Samuel and Scialom, David},
  journal={Physical Review D},
  volume={57},
  number={10},
  pages={6065},
  year={1998},
  publisher={APS}
}

@article{ito2016mhz,
  title={MHz gravitational waves from short-term anisotropic inflation},
  author={Ito, Asuka and Soda, Jiro},
  journal={Journal of Cosmology and Astroparticle Physics},
  volume={2016},
  number={04},
  pages={035},
  year={2016},
  publisher={IOP Publishing}
}

@article{franco2017tensor,
  title={Tensor perturbations in anisotropically curved cosmologies},
  author={Franco, Felipe O and Pereira, Thiago S},
  journal={Journal of Cosmology and Astroparticle Physics},
  volume={2017},
  number={11},
  pages={022},
  year={2017},
  publisher={IOP Publishing}
}

@article{hertzberg2024constraints,
  title={Constraints on an anisotropic universe},
  author={Hertzberg, Mark P and Loeb, Abraham},
  journal={Physical Review D},
  volume={109},
  number={8},
  pages={083538},
  year={2024},
  publisher={APS}
}

@article{maleknejad2012revisiting,
  title={Revisiting cosmic no-hair theorem for inflationary settings},
  author={Maleknejad, A and Sheikh-Jabbari, MM},
  journal={Physical Review D—Particles, Fields, Gravitation, and Cosmology},
  volume={85},
  number={12},
  pages={123508},
  year={2012},
  publisher={APS}
}

@article{soda2012statistical,
  title={Statistical anisotropy from anisotropic inflation},
  author={Soda, Jiro},
  journal={Classical and Quantum Gravity},
  volume={29},
  number={8},
  pages={083001},
  year={2012},
  publisher={IOP Publishing}
}

@article{bennett1994cosmic,
  title={Cosmic temperature fluctuations from two years of COBE DMR observations},
  author={Bennett, C and Kogut, A and Hinshaw, G and Banday, A and Wright, E and Gorski, K and Wilkinson, D and Weiss, R and Smoot, G and Meyer, S and others},
  journal={arXiv preprint astro-ph/9401012},
  year={1994}
}

@inproceedings{smoot1999cobe,
  title={COBE observations and results},
  author={Smoot, George F},
  booktitle={AIP Conference Proceedings CONF-981098},
  volume={476},
  number={1},
  pages={1--10},
  year={1999},
  organization={American Institute of Physics}
}

@article{montroy2003measuring,
  title={Measuring CMB polarization with Boomerang},
  author={Montroy, Tom and Ade, Peter AR and Balbi, A and Bock, JJ and Bond, JR and Borrill, J and Boscaleri, A and Cabella, Paolo and Contaldi, CR and Crill, BP and others},
  journal={New Astronomy Reviews},
  volume={47},
  number={11-12},
  pages={1057--1065},
  year={2003},
  publisher={Elsevier}
}

@article{gurzadyan2003ellipticity,
  title={Ellipticity analysis of the BOOMERanG CMB maps},
  author={Gurzadyan, VG and Ade, Peter AR and De BERNARDIS, Paolo and Bianco, Carlo Luciano and Bock, JJ and Boscaleri, A and Crill, BP and De TROIA, Grazia and Ganga, K and Giacometti, M and others},
  journal={International Journal of Modern Physics D},
  volume={12},
  number={10},
  pages={1859--1873},
  year={2003},
  publisher={World Scientific}
}

@article{carlstrom2003status,
  title={Status of CMB polarization measurements from DASI and other experiments},
  author={Carlstrom, JE and Kovac, J and Leitch, EM and Pryke, C},
  journal={New Astronomy Reviews},
  volume={47},
  number={11-12},
  pages={953--966},
  year={2003},
  publisher={Elsevier}
}

@article{hou2014constraints,
  title={Constraints on cosmology from the cosmic microwave background power spectrum of the 2500 deg2 SPT-SZ survey},
  author={Hou, Z and Reichardt, CL and Story, KT and Follin, B and Keisler, R and Aird, KA and Benson, BA and Bleem, LE and Carlstrom, JE and Chang, CL and others},
  journal={The Astrophysical Journal},
  volume={782},
  number={2},
  pages={74},
  year={2014},
  publisher={IOP Publishing}
}

@article{leitch2002measuring,
  title={Measuring polarization with DASI},
  author={Leitch, Erik M and Kovac, JM and Pryke, C and Reddall, B and Sandberg, ES and Dragovan, M and Carlstrom, JE and Halverson, NW and Holzapfel, WL},
  journal={arXiv preprint astro-ph/0209476},
  year={2002}
}

@article{doroshkevich2004large,
  title={Large scale structure in the SDSS galaxy survey},
  author={Doroshkevich, Andrei and Tucker, DL and Allam, S and Way, MJ},
  journal={Astronomy \& Astrophysics},
  volume={418},
  number={1},
  pages={7--23},
  year={2004},
  publisher={EDP Sciences}
}

@article{to2025dark,
  title={Dark energy survey: Modeling strategy for multiprobe cluster cosmology and validation for the full six-year dataset},
  author={To, Chun-Hao and Krause, Elisabeth and Chang, Chihway and Wu, Hao-Yi and Wechsler, Risa H and Rozo, Eduardo and Weinberg, David H and Anbajagane, D and Avila, S and Blazek, J and others},
  journal={Physical Review D},
  volume={112},
  number={6},
  pages={063537},
  year={2025},
  publisher={APS}
}

@article{labini2025large,
  title={Large-Scale Galaxy Correlations from the DESI First Data Release},
  author={Labini, Francesco Sylos and Antal, Tibor},
  journal={arXiv preprint arXiv:2511.21585},
  year={2025}
}

@article{aiola2020atacama,
  title={The Atacama Cosmology Telescope: DR4 maps and cosmological parameters},
  author={Aiola, Simone and Calabrese, Erminia and Maurin, Lo{\"\i}c and Naess, Sigurd and Schmitt, Benjamin L and Abitbol, Maximilian H and Addison, Graeme E and Ade, Peter AR and Alonso, David and Amiri, Mandana and others},
  journal={Journal of Cosmology and Astroparticle Physics},
  volume={2020},
  number={12},
  pages={047},
  year={2020},
  publisher={IOP Publishing}
}

@article{riess1998observational,
  title={Observational evidence from supernovae for an accelerating universe and a cosmological constant},
  author={Riess, Adam G and Filippenko, Alexei V and Challis, Peter and Clocchiatti, Alejandro and Diercks, Alan and Garnavich, Peter M and Gilliland, Ron L and Hogan, Craig J and Jha, Saurabh and Kirshner, Robert P and others},
  journal={The astronomical journal},
  volume={116},
  number={3},
  pages={1009},
  year={1998},
  publisher={IOP Publishing}
}

@article{frieman2008dark,
  title={Dark energy and the accelerating universe},
  author={Frieman, Joshua A and Turner, Michael S and Huterer, Dragan},
  journal={Annu. Rev. Astron. Astrophys.},
  volume={46},
  number={1},
  pages={385--432},
  year={2008},
  publisher={Annual Reviews}
}

@article{cooke2014precision,
  title={Precision measures of the primordial abundance of deuterium},
  author={Cooke, Ryan J and Pettini, Max and Jorgenson, Regina A and Murphy, Michael T and Steidel, Charles C},
  journal={The Astrophysical Journal},
  volume={781},
  number={1},
  pages={31},
  year={2014},
  publisher={IOP Publishing}
}

@article{peebles2025status,
  title={Status of the $\Lambda$CDM theory: supporting evidence and anomalies},
  author={Peebles, Phillip James E},
  journal={Philosophical Transactions A},
  volume={383},
  number={2290},
  pages={20240021},
  year={2025},
  publisher={The Royal Society}
}

@article{akarsu2024lambdacdm,
  title={$\Lambda$CDM tensions: localising missing physics through consistency checks},
  author={Akarsu, {\"O}zg{\"u}r and {\'O} Colg{\'a}in, Eoin and Sen, Anjan A and Sheikh-Jabbari, MM},
  journal={Universe},
  volume={10},
  number={8},
  pages={305},
  year={2024},
  publisher={MDPI}
}

@article{de2004significance,
  title={Significance of the largest scale CMB fluctuations in WMAP},
  author={de Oliveira-Costa, Angelica and Tegmark, Max and Zaldarriaga, Matias and Hamilton, Andrew},
  journal={Physical Review D},
  volume={69},
  number={6},
  pages={063516},
  year={2004},
  publisher={APS}
}

@article{copi2004multipole,
  title={Multipole vectors: A new representation of the CMB sky and evidence for statistical anisotropy or non-Gaussianity at 2$\leqslant$ l$\leqslant$ 8},
  author={Copi, Craig J and Huterer, Dragan and Starkman, Glenn D},
  journal={Physical Review D},
  volume={70},
  number={4},
  pages={043515},
  year={2004},
  publisher={APS}
}

@article{koivisto2011possibility,
  title={Possibility of anisotropic curvature in cosmology},
  author={Koivisto, Tomi S and Mota, David F and Quartin, Miguel and Zlosnik, Tom G},
  journal={Physical Review D—Particles, Fields, Gravitation, and Cosmology},
  volume={83},
  number={2},
  pages={023509},
  year={2011},
  publisher={APS}
}

@article{graham2010observing,
  title={Observing the dimensionality of our parent vacuum},
  author={Graham, Peter W and Harnik, Roni and Rajendran, Surjeet},
  journal={Physical Review D—Particles, Fields, Gravitation, and Cosmology},
  volume={82},
  number={6},
  pages={063524},
  year={2010},
  publisher={APS}
}

@article{smith2025cosmological,
  title={Cosmological constraints on anisotropic Thurston geometries},
  author={Smith, Ananda F and Copi, Craig J and Starkman, Glenn D},
  journal={Journal of Cosmology and Astroparticle Physics},
  volume={2025},
  number={01},
  pages={005},
  year={2025},
  publisher={IOP Publishing}
}

@book{weeks2001shape,
  title={The shape of space},
  author={Weeks, Jeffrey R},
  year={2001},
  publisher={CRC press}
}

@article{liu2009improved,
  title={Improved CMB map from WMAP data},
  author={Liu, Hao and Li, Ti-Pei},
  journal={arXiv preprint arXiv:0907.2731},
  year={2009}
}

@article{souradeep2006angular,
  title={Angular power spectrum of CMB anisotropy from WMAP},
  author={Souradeep, Tarun and Saha, Rajib and Jain, Pankaj},
  journal={New Astronomy Reviews},
  volume={50},
  number={11-12},
  pages={854--860},
  year={2006},
  publisher={Elsevier}
}

@article{tegmark2003high,
  title={High resolution foreground cleaned CMB map from WMAP},
  author={Tegmark, Max and de Oliveira-Costa, Angelica and Hamilton, Andrew JS},
  journal={Physical Review D},
  volume={68},
  number={12},
  pages={123523},
  year={2003},
  publisher={APS}
}

@article{schwarz2016cmb,
  title={CMB anomalies after Planck},
  author={Schwarz, Dominik J and Copi, Craig J and Huterer, Dragan and Starkman, Glenn D},
  journal={Classical and Quantum Gravity},
  volume={33},
  number={18},
  pages={184001},
  year={2016},
  publisher={IOP Publishing}
}

@article{lamarre2003planck,
  title={The Planck High Frequency Instrument, a third generation CMB experiment, and a full sky submillimeter survey},
  author={Lamarre, Jean Michael and Puget, JL and Bouchet, Freddy and Ade, Peter AR and Benoit, Ang{\'e}lique and Bernard, Jean-Paul and Bock, J and De Bernardis, Paolo and Charra, J and Couchot, F and others},
  journal={New Astronomy Reviews},
  volume={47},
  number={11-12},
  pages={1017--1024},
  year={2003},
  publisher={Elsevier}
}

@article{scott1995microwave,
  title={From microwave anisotropies to cosmology},
  author={Scott, Douglas and Silk, Joseph and White, Martin},
  journal={Science},
  volume={268},
  number={5212},
  pages={829--835},
  year={1995},
  publisher={American Association for the Advancement of Science}
}

@article{sugiyama1994cosmic,
  title={Cosmic background anisotropies in CDM cosmology},
  author={Sugiyama, Naoshi},
  journal={arXiv preprint astro-ph/9412025},
  year={1994}
}

@article{wojtak2013orbital,
  title={Orbital anisotropy in cosmological haloes revisited},
  author={Wojtak, Rados{\l}aw and Gottl{\"o}ber, Stefan and Klypin, Anatoly},
  journal={Monthly Notices of the Royal Astronomical Society},
  volume={434},
  number={2},
  pages={1576--1585},
  year={2013},
  publisher={The Royal Astronomical Society}
}

@article{bunn1996anisotropic,
  title={How anisotropic is our universe?},
  author={Bunn, Emory F and Ferreira, Pedro G and Silk, Joseph},
  journal={Physical Review Letters},
  volume={77},
  number={14},
  pages={2883},
  year={1996},
  publisher={APS}
}

@article{bartlett1999standard,
  title={The standard cosmological model and CMB anisotropies},
  author={Bartlett, James G},
  journal={New Astronomy Reviews},
  volume={43},
  number={2-4},
  pages={83--109},
  year={1999},
  publisher={Elsevier}
}

@article{alpher1948evolution,
  title={Evolution of the Universe},
  author={Alpher, Ralph A and Herman, Robert},
  journal={Nature},
  volume={162},
  number={4124},
  pages={774--775},
  year={1948},
  publisher={Nature Publishing Group UK London}
}

@article{peebles1966primordial,
  title={Primordial helium abundance and the primordial fireball. II},
  author={Peebles, P James E},
  journal={Astrophysical Journal, vol. 146, p. 542},
  volume={146},
  pages={542},
  year={1966}
}

@article{wagoner1967synthesis,
  title={On the synthesis of elements at very high temperatures},
  author={Wagoner, Robert V and Fowler, William A and Hoyle, Fred},
  journal={Astrophysical Journal, vol. 148, p. 3},
  volume={148},
  pages={3},
  year={1967}
}

@article{reeves1974origin,
  title={On the origin of the light elements},
  author={Reeves, R},
  journal={In: Annual review of astronomy and astrophysics. Volume 12.(A75-13476 03-90) Palo Alto, Calif., Annual Reviews, Inc., 1974, p. 437-469.},
  volume={12},
  pages={437--469},
  year={1974}
}

@article{esmailzadeh1991primordial,
  title={Primordial nucleosynthesis without a computer},
  author={Esmailzadeh, Rahim and Starkman, Glenn D and Dimopoulos, Savas},
  journal={Astrophysical Journal, Part 1 (ISSN 0004-637X), vol. 378, Sept. 10, 1991, p. 504-518. Research supported by Keck Foundation and New Jersey High Technology Program.},
  volume={378},
  pages={504--518},
  year={1991}
}

@article{cyburt2016big,
  title={Big bang nucleosynthesis: Present status},
  author={Cyburt, Richard H and Fields, Brian D and Olive, Keith A and Yeh, Tsung-Han},
  journal={Reviews of Modern Physics},
  volume={88},
  number={1},
  pages={015004},
  year={2016},
  publisher={APS}
}

@article{copi2010large,
  title={Large-Angle Anomalies in the CMB},
  author={Copi, Craig J and Huterer, Dragan and Schwarz, Dominik J and Starkman, Glenn D},
  journal={Advances in Astronomy},
  volume={2010},
  number={1},
  pages={847541},
  year={2010},
  publisher={Wiley Online Library}
}

@article{bunn2010large,
  title={Large-angle anomalies in the microwave background},
  author={Bunn, Emory F},
  journal={arXiv preprint arXiv:1006.2084},
  year={2010}
}

@article{smoot1992structure,
  title={Structure in the COBE differential microwave radiometer first-year maps},
  author={Smoot, George F and Bennett, Charles L and Kogut, Alan and Wright, Edward L and Aymon, John and Boggess, Nancy W and Cheng, Edward S and De Amici, Giovanni and Gulkis, Steven and Hauser, Michael G and others},
  journal={Astrophysical Journal, Part 2-Letters (ISSN 0004-637X), vol. 396, no. 1, Sept. 1, 1992, p. L1-L5. Research supported by NASA.},
  volume={396},
  pages={L1--L5},
  year={1992}
}

@article{bennett2003microwave,
  title={The microwave anisotropy probe mission},
  author={Bennett, Charles L and Bay, M and Halpern, M and Hinshaw, G and Jackson, C and Jarosik, N and Kogut, A and Limon, M and Meyer, SS and Page, L and others},
  journal={The Astrophysical Journal},
  volume={583},
  number={1},
  pages={1--23},
  year={2003}
}

@article{ade2016planck,
  title={Planck 2015 results-xiii. cosmological parameters},
  author={Ade, Peter AR and Aghanim, Nabila and Arnaud, M and Ashdown, Mark and Aumont, Jea and Baccigalupi, Carlo and Banday, AJ and Barreiro, RB and Bartlett, JG and Bartolo, Nicola and others},
  journal={Astronomy \& Astrophysics},
  volume={594},
  pages={A13},
  year={2016},
  publisher={EDP sciences}
}

@article{n2020planck,
  title={Planck 2018 results. I. Overview and the cosmological legacy of Planck},
  author={N Aghanim, YA and Arroja, F and Aumont, J and Baccigalupi, C and Ballardini, M and Banday, AJ and RB Barreiro, NB and Basak, S and Battye, R and Benabed, K and others},
  journal={Astronomy \& Astrophysics},
  volume={641},
  year={2020},
  publisher={EDP Sciences}
}

@article{copi2007uncorrelated,
  title={Uncorrelated universe: Statistical anisotropy and the vanishing angular correlation function<? format?> in WMAP years 1--3},
  author={Copi, Craig J and Huterer, Dragan and Schwarz, Dominik J and Starkman, Glenn D},
  journal={Physical Review D—Particles, Fields, Gravitation, and Cosmology},
  volume={75},
  number={2},
  pages={023507},
  year={2007},
  publisher={APS}
}

@article{hansen2009power,
  title={Power asymmetry in cosmic microwave background fluctuations from full sky to sub-degree scales: is the universe isotropic?},
  author={Hansen, FK and Banday, AJ and Gorski, KM and Eriksen, HK and Lilje, PB},
  journal={The Astrophysical Journal},
  volume={704},
  number={2},
  pages={1448--1458},
  year={2009},
  publisher={The American Astronomical Society}
}
\end{document}